\documentclass[fleqn,usenatbib]{mnras}

\usepackage{newtxtext,newtxmath}
\usepackage[T1]{fontenc}

\DeclareRobustCommand{\VAN}[3]{#2}
\let\VANthebibliography\thebibliography
\def\thebibliography{\DeclareRobustCommand{\VAN}[3]{##3}\VANthebibliography}

\usepackage{graphicx}	% Including figure files
\usepackage{amsmath}	% Advanced maths commands
\usepackage{float}
\usepackage{subcaption}

\usepackage[usenames]{color}
\usepackage{booktabs}
\usepackage{xcolor}
\usepackage{verbatim}
\usepackage{caption}
\usepackage{subcaption}
\usepackage{xltabular} 
\usepackage{ragged2e} % Para \RaggedRight
\usepackage{array}  % for custom column widths
\usepackage[normalem]{ulem}
\usepackage{hyperref}

\newcolumntype{P}[1]
{>{\centering\arraybackslash}p{#1}}
\title[A mock DSFG catalogue for mm surveys]{A mock redshift catalogue of the dusty star-forming galaxy population with intrinsic clustering and lensing for deep millimetre surveys}

\author[A. Nava-Moreno]{    
    Araceli Nava-Moreno,$^{1}$\thanks{E-mail: aracelinavam@gmail.com}
    Alfredo Montaña,$^{1}$
    Itziar Aretxaga,$^{1}$
    Aldo Rodríguez-Puebla,$^{2}$
    \newauthor
    Vladimir Avila-Reese$^{2}$, Edgar Peralta$^{1}$
    \\
    $^{1}$Instituto Nacional de Astrofísica, Óptica y Electrónica (INAOE), Luis Enrique Erro 1, Sta. Ma. Tonantzintla, 72840, Puebla, México\\
    $^{2}$Universidad Nacional Aut\'onoma de M\'exico, Instituto de Astronom\'ia, A. P. 70-264, 04510, Ciudad de M\'exico, M\'exico
    }
    
\date{Accepted XXX. Received YYY; in original form ZZZ}

\pubyear{2024}

\begin{document}
\label{firstpage}
\pagerange{\pageref{firstpage}--\pageref{lastpage}}
\maketitle

%----------------- Abstract of the paper -------------------------
\begin{abstract}
We present a new cosmologically motivated mock redshift survey of the Dusty Star-Forming Galaxy population. Our mock survey is based on the Bolshoi--\textit{Planck} dark-matter halo simulation and covers an area of 5.3 deg$^{2}$. Using a semi-empirical approach, we generate a light cone and populate the dark-matter haloes with galaxies. Infrared properties are assigned to the galaxies based on theoretical and empirical relations from the literature. Additionally, background galaxies are gravitationally lensed by dark-matter haloes along the line-of-sight assuming a point-mass model approximation. We characterize the mock survey by measuring the star formation rate density, integrated number counts, redshift distribution, and infrared luminosity function. When compared with single-dish and interferometric observations, the predictions from our mock survey closely follow the compiled results from the literature. We have also directed this study towards characterizing one of the extragalactic legacy surveys to be observed with the TolTEC camera at the Large Millimeter Telescope: the 0.8 sq. degree Ultra Deep Survey, with expected depths of 0.025, 0.018 and 0.012 mJy beam$^{-1}$ at 1.1, 1.4 and 2.0 mm. Exploiting the clustering information in our mock survey, we investigate its impact on the effect of flux boosting by the fainter population of dusty galaxies, finding that clustering can increase the median boosting by 0.5\,per\,cent at 1.1 mm, 0.8\,per\,cent at 1.4 mm and, 2.0\,per\,cent at 2.0 mm, and with higher dispersion.
\end{abstract}
%----------------------------------------------------------------

% Select between one and six entries from the list of approved keywords.
% Don't make up new ones.We investigate the impact of including clustering information on the measurement of boosted flux densities

%%%%%%%%%%%%%%%%%%%%%%%%%%%%%%%%%%%%%%%%%%%%%%%%%%%%%%%%%%
%%%%%%%%%%%%%%%%%     BODY OF PAPER     %%%%%%%%%%%%%%%%%%
%%%%%%%%%%%%%%%%%%%%%%%%%%%%%%%%%%%%%%%%%%%%%%%%%%%%%%%%%%

\section{Introduction}\label{intro}

A key piece to unravel the formation and evolution of galaxies is the study of the dusty star-forming galaxy (DSFG) population. These galaxies are highly obscured by dust, preventing the estimation of total star formation rates (SFR) at UV-optical wavelengths. However, at submillimetre wavelengths, we can observe the emission of dust grains heated by stars. The camera SCUBA in the James Clerk Maxwell Telescope (JCMT) discovered a subset of DSFG known as submillimeter galaxies  \citep[SMG, e.g.,][]{smail+1997, barger+1998, hughes+1998}. These galaxies are found mainly at high redshifts, with a peak at $z \sim 2-3$ \citep[][]{chapman+2003, aretxaga+2003, chapman+2005, aretxaga+2005, stach+2018, franco+2020, gomez-guijarro+2022}, and they exhibit high infrared luminosities \citep[$L_{\rm{IR}}>10^{12}$L$_{\sun}$,][]{smail+1997, hughes+1998, magnelli+2013, gruppioni+2020, casey+2013}, corresponding to extreme SFRs of hundreds and even thousands of M$_{\sun}$yr$^{-1}$ \citep[e.g.][]{smail+1997, hughes+1998, barger+2014, da_cunha+2015}. For a review on this population see \citep{casey+2014a}. The SFR density (SFRD) is highly obscured at $z\lesssim 3$, with a contribution from DSFGs reaching approximately fifty per cent at $z\approx 3$ \citep{bourne+2017, Dudzeviciute+2020}. However, at $z\gtrsim 4$ their contribution is still not well constrained since the samples of DSFGs at these high redshifts remain incomplete \citep[e.g.][]{casey+2014a, gruppioni+2020, khusanova+2020, zavala+2021, fudamoto+2021, algera+2022}. Measuring reliable photometric or spectroscopic redshifts of DSFGs often requires counterpart associations across UV to mid-infrared wavelengths. However, detecting high$-z$ galaxies at these wavelengths becomes challenging due to their flux densities rapidly dropping with redshift.
Additionally, the relatively poor angular resolution and sensitivity of the single-dish (sub-)mm telescopes further complicate the association between the DSFGs and their UV/optical counterparts. The single-dish telescopes observe only the most luminous galaxies in large areas, e.g., SCUBA/SCUBA-2 on the JCMT \citep{geach+2017}, MAMBO on IRAM \citep{bertoldi+2007}, SPT \citep{vieira+2010}, and AzTEC on ASTE \citep{aretxaga+2011}. Interferometric telescopes as ALMA have shown that a considerable fraction of these luminous galaxies are actually the result of blending fainter sources within the larger beam of single-dish telescopes \citep[e.g.][]{karim+2013, simpson+2015, stach+2018}: some cases being the result of projected sources in the line-of-sight, while others have been confirmed to be physically associated galaxies potentially tracing over-dense regions of the Universe.  
Due to their high angular resolution and sensitivity, interferometers allow more detailed studies of DSFGs. However, compared to single-dish observations, interferometric surveys are limited to mapping relatively smaller areas of the sky \citep[hundreds of square arcminutes, e.g. ][]{Lindner+2011, fujimoto+2016, dunlop+2017, franco+2018}. To obtain a complete sample of DSFGs on both small and large scales and to understand their role not only in the SFRD \citep{madau_dickinson2014, dunlop+2017, magnelli+2019, gruppioni+2020} but also in their physical evolution, there is a need for a link between single-dish and interferometric telescopes.

The new camera TolTEC\footnote{http://toltec.astro.umass.edu/} \citep{wilson+2020} installed in the 50m Large Millimeter Telescope Alfonso Serrano \citep[LMT\footnote{http://lmtgtm.org/},][]{hughes+2020} observes at three different wavelengths simultaneously (1.1, 1.4 and 2.0 mm), with angular resolutions of 5.0, 6.3 and 9.5 arcsec. TolTEC will image a series of Extragalactic Legacy Surveys, two of which have already been defined: the Ultra Deep Survey (UDS) and the Large Scale Structure Survey (LSS) \citep[][]{montana+2019}.

The UDS survey will be sensitive to typical star forming galaxies ($\rm{SFR}>10$ M$_{\sun}$ yr$^{-1}$ and $L_{\rm{IR}}>10^{11}$ L$_{\sun}$) with $1\sigma$ rms depths of 0.025, 0.018 and 0.012 mJy beam$^{-1}$ at 1.1, 1.4 and 2.0 mm. The goal is to study the history of stellar mass build-up and metal production in massive galaxies during the last 13 Gyr of cosmic time \citep{montana+2019}. The total survey area will cover $\sim$ 0.8 deg$^{2}$, observing the CANDELS fields in UDS, GOODS--S, and a 0.5 sq. degree area in COSMOS. On the other hand, the LSS survey will cover a larger area of the sky to investigate the spatial distribution of the star-forming galaxies through the cosmic web. It will study galaxies with $\rm{SFR}>100$ M$_{\sun}$ yr$^{-1}$ and $L_{\rm{IR}}>10^{12}$ L$_{\sun}$ at 1.1, 1.4 and 2.0 mm with depths of 0.25, 0.18 and 0.12 mJy beam${-1}$ at $1\sigma$ in rms. The observed areas will be of 3.5 to 10 deg$^{2}$ for target individual fields and a total of 40-60 deg$^{2}$ for the full survey \citep{montana+2019}. These two extragalactic legacy surveys will close the gap between single-dish and interferometric observations.

Works based on N-body cosmological simulations and different methods to populate galaxies within the dark matter halos \citep[e.g.,][]{cowley+2014, bethermin+2017, popping+2020} or in hydrodynamics simulations \citep[e.g.,][]{lovell+2021}, have attempted to recreate the IR properties of the DSFGs and consequently reproduce the integrated number counts. These simulations allow us to understand better the DSFG population, and can be used to study the impact of observational effects (flux boosting, confusion noise, source blending, completeness, cosmic variance, etc.) on the detection of galaxies and different observables, e.g., the measured number counts \citep{cowley+2014, bethermin+2017} and angular correlation functions \citep{cowley+2016,cowley+2017}. The flux boosting effect is present when unresolved neighbouring galaxies fall within the same telescope beam and increase the measured flux density. 
The flux boosting corrections to individual galaxies are commonly estimated through simulations that randomly distribute the galaxies in the observed area \citep[e.g.,][]{coppin+2005, coppin+2006,geach+2017, zavala+2017, chen+2022}. These latter simulations do not consider the clustering of the DSFGs. However including clustering potentially has an impact on the flux boosting effects since it depends on the likelihood of having nearby galaxies.

In this paper, we present a new mock redshift survey of the DSFG population. This mock survey incorporates clustering properties and the gravitational lensing magnification caused by the dark-matter haloes in the line-of-sight on background galaxies, as it is built on a dark-matter cosmological simulation. As discussed above, previous mock surveys have overlooked both of these effects. We take advantage of this new mock redshift survey to characterize the TolTEC/UDS extragalactic legacy survey and to explore the impact that the clustering of DSFGs might have in the determination and correction of flux density boosting effects. 

In section \ref{sec2:simulation} we describe our procedure to generate our mock redshift survey of the DSFG population. In section \ref{sec3:sim_characterization} we characterize our mock redshift survey and compare it with the SFRD, redshift distribution, number counts, and the luminosity function of DSFGs. In Section \ref{sec3:pred_toltec_areas} we show our predictions for the UDS extragalactic legacy survey of TolTEC and we discuss the different observational effects present in this survey. In section \ref{sec4:boosting_clustering} we analyze the impact in the corrections of flux boosting when the clustering of galaxies is taken into account. Finally, we present our conclusions in the Section \ref{sec5:discussion_conclution}. 

In this work we adopt a flat $\Lambda$CDM cosmology with: $\Omega_{m}$ = 0.307, $\Omega_{\Lambda}$ = 0.693 and $h$ = 0.678 \citep{planck+2016}.

%---------------SECTION 2-------------

\section{Simulation and mock redshift survey}\label{sec2:simulation}   

 During the past years, the Simulated Infrared Dusty Extragalactic Sky \citep[SIDES,][]{bethermin+2017} has been the largest (2 deg$^{2}$) publicly available simulation of the DSFG population. SIDES accurately reproduces the FIR-mm number counts of the DSFGs and, since it is based on the Bolshoi-\textit{Planck} cosmological simulation, it therefore includes their clustering properties. SIDES has been widely used to compare and analyze observations of the DSFG population \citep{bethermin+2020,chen+2022,bing+2023,traina+2024} and has recently been updated to predict their emission line properties \cite[CONCERTO,][]{bethermin+2022}. The 2 deg$^{2}$ covered by SIDES, however, restricts its use to the analysis of relatively small area surveys, limits the predictions of cosmic variance effects, and misses the brightest and more extreme SMGs. The larger areas of future millimetre wavelength surveys, particularly those to be conducted with the TolTEC camera, has motivated the development of the new larger area (5.3 deg$^{2}$) mock redshift survey presented in this work. 

In this section we describe the procedure followed to build our mock redshift survey of the DSFG population. First we describe the construction of a lightcone of dark matter haloes and how these haloes are populated with galaxies. We latter explain how star-forming galaxies are identified within the mock redshift survey, and the process followed to assign their infrared properties. Finally, we estimate gravitational lensing magnifications and the observed flux densities for each galaxy. In Appendix \ref{tab:appendix_A} we highlight the main differences between our model (described in \S\ref{subsec2:lightcone} - \S\ref{subsec2:grav_lens}) and that from the SIDES simulation.

\subsection{Lightcone} \label{subsec2:lightcone}

We use the Bolshoi-\textit{Planck} cosmological simulation \citep{klypin+2016, RP+2016b}, which has a side length of 250$h^{-1}$ Mpc, with 2048$^{3}$ dark matter particles and a mass resolution of 1.55$\times 10^{8}$ $h^{-1}$M$_{\sun}$. The periodic conditions of the Bolshoi-\textit{Planck} simulation allows us to build a lightcone, covering a wide range of redshifts ($z \lesssim 7$) by repeating the volume of the box in the $z$ direction as many times as necessary. The projection of the Bolshoi-\textit{Planck} box results in a 5.3 deg$^{2}$ area. This size is large enough to generate multiple synthetic observations of the maps expected for the UDS. The maximum number necessary to cover a square area, with a side length $\theta$, is given by:

\begin{equation}
    N_{\rm{repetitions}} = \rm{Integer}\left ( \frac{360^{\circ}}{2\pi\theta}+1 \right ),
    \label{eq:repetitions}
\end{equation} 

\noindent where Integer(x) rounds the value of $x$ to the nearest integer. Thus, for a square area of $5.3\; \rm{deg}^{2}$ $N_{\rm{repetitions}}=26$, the corresponding comoving volume covered by the lightcone is $V \approx 1.15 \times 10^{5}  \;h^{-3}$Gpc$^{3}$ up to redshift 7.

Due to the box repetitions, the light-cone projection on the sky can present false radial patterns. To avoid this, we include random displacements of the boxes no greater than $10\%$ of the length of the box in one of the axis as well as several random rotations and permutations of each box. A visual inspection of the lightcone projection shows these are satisfactory solutions. To establish a connection between the galaxies and their host dark-matter haloes, we use an updated version to the semi-empirical galaxy-halo connection developed by \citet{RP+2016a, RP+2017}. The general idea is to match the halo number density and the galaxy number density through a particular combination of their properties while at the same time modelling the mass assembly and star formation history of the galaxies. Here we use the maximum circular velocity attained along the main progenitor branch of the halo merger trees ($V_\text{peak}$) and the stellar mass ($M_{\star}$) of the galaxies to make the connection. The above is well known that reproduces the auto-correlation functions of galaxies as a function of stellar mass \citep[see e.g.,][]{Reddick+2013,Dragomir+2018,Calette+2021} and when dividing into star-forming and quiescent galaxies \citep{Kakos+2024}.

SFRs are determined following the growth in mass of the host haloes and their associated galaxies through the merger trees of the Bolshoi-\textit{Planck} simulation, dividing the stellar mass growth into two components: 1) $in-situ$ star formation and 2) $ex-situ$ star formation due to the merger of smaller galaxies. The history of stellar mass growth is calibrated 
to reproduce the evolution of the stellar mass function of all galaxies from $z \sim 0.1$ to $z \sim 10$, the fraction of quiescent galaxies from $z\sim0.1$ to $\sim4$, and the mean of the star-forming main sequence of galaxies from $z\sim0.1$ to $z\sim8$ (Rodríguez-Puebla et al. in prep.). In this way, each dark-matter halo and subhalo in our lightcone hosts a galaxy with a defined $M_{\star}$ and SFR.

It is important to note that the semi-empirical approach does not introduce any modelling of astrophysical processes (as in the case of semi-analytical models) or recipes to physically link halo evolution with galaxy evolution. The approach is based on a continuous statistical matching of the simulated halo and observed galaxy distributions over a wide redshift range. As a result, the average stellar mass growth (in-situ and ex-situ) and the SFR history of galaxies bound to their dark matter haloes in the simulation are predicted. However, despite the empirical nature of the approach, it contains some underlying assumptions, for example, that the IMF (Initial Mass Function) is universal. Remarkably, after introducing observed systematic and statistical uncertainties in the galaxy-halo connection modelling, the integration in time of the SFR histories of galaxies in the mock catalogues is fully consistent with the evolution of the galaxy stellar mass function, not being necessary to invoke a $z$- and mass-dependent IMF.

\begin{figure}
\includegraphics[width=\columnwidth]{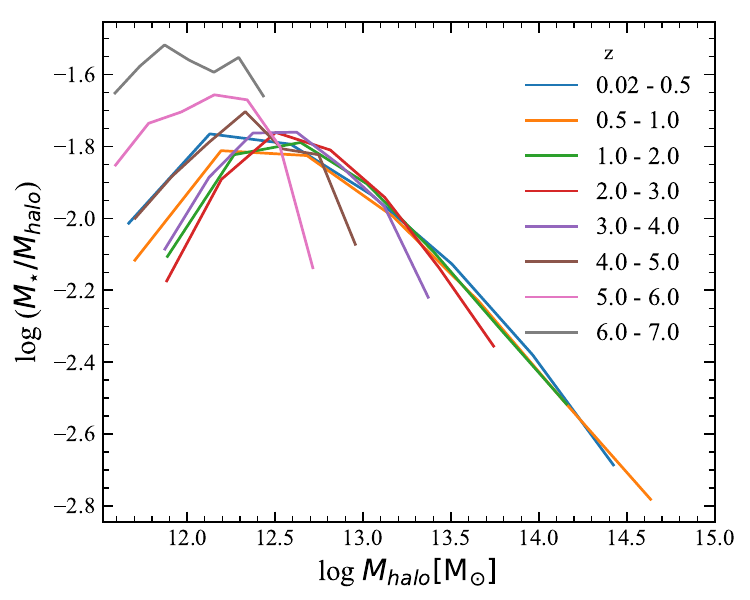}
    \caption{Evolution of the stellar-to-halo mass ratio since $z \sim 0.02$ to $z \sim 6.5$ from our mock redshift survey \citep[Rodriguez-Puebla et al. in prep, which represents an update to][]{RP+2017}. Halo masses refer to virial masses, which at $z \sim 0$ are 333 times the mean matter background density and at high $z$ approaches to 178. We note that stellar masses are equivalent to observed stellar masses since they include random errors as described in \citet{RP+2017}.}
    \label{fig:ratio_ms_mh}
\end{figure}

Figure \ref{fig:ratio_ms_mh} shows the predicted stellar-to-halo mass ratio at different redshifts, where this ratio quantifies the time-integrated efficiency of star formation in halos.  As seen, the stellar-to-halo mass ratio is on average independent of redshift. In detail, however, it peaks around  $M_{{\rm halo}}\sim 10^{12.0}$ M$_{\odot}$ at $z \sim 0.25$, increases slightly with $z$ up to $M_{{\rm halo}}\sim 10^{12.5}$ at $z \sim 3$, and then it decreases to smaller masses at even higher redshifts. At masses below the peak, the role of star formation feedback is supposed to be crucial, while at higher masses, AGN feedback and virial-shocked gas cooling are supposed to be  relevant. 

    \subsection{Selection of Star-Forming Galaxies}\label{subsec2:sf_selec}
    
For the purposes of this work, it is necessary to make a selection of galaxies that are actively forming stars. To achieve this, we initially calculate the specific star formation rate (sSFR = SFR/$M_{\star}$, a measure of the star formation per unity mass) for each galaxy within our mock catalogue. Subsequently, we apply the criterion of \citet{pacifici+2016} to preselect star-forming galaxies. This criterion parametrizes the redshift evolution of the sSFR threshold between star-forming and quiescent galaxies as:

\begin{equation}
{\rm{sSFR_{min}}}=0.2/t_{\rm{u}}(z),
\label{ec:pacifici}
\end{equation}

\noindent where $t_{u}(z)$ is the age of the Universe in Gyr at redshift $z$ and ${\rm{sSFR_{min}}}$ is in Gyr$^{-1}$. We calculate ${\rm{sSFR_{min}}}$ for each galaxy in our mock catalogue and compare this value with its intrinsic sSFR. If $\rm{sSFR}>\rm{sSFR_{min}}$ the object is selected as a candidate star-forming galaxy. We then group our mock galaxies in redshift and stellar mass bins, calculate the mean and standard deviation of the sSFR per bin, and make a linear fit to these values. The selected star-forming galaxies will be those that are above the fit minus 0.5 dex. For a better selection, the mean values are recalculated using this last selected subset of galaxies, and a new linear fit is performed. This sigma clipping process continues until the linear fit converges and no longer exhibits significant variation ($\Delta < 0.05$). The last fit defines the main sequence of our star-forming galaxies. 

\begin{figure*}
    \centering
    \includegraphics[width=\textwidth]{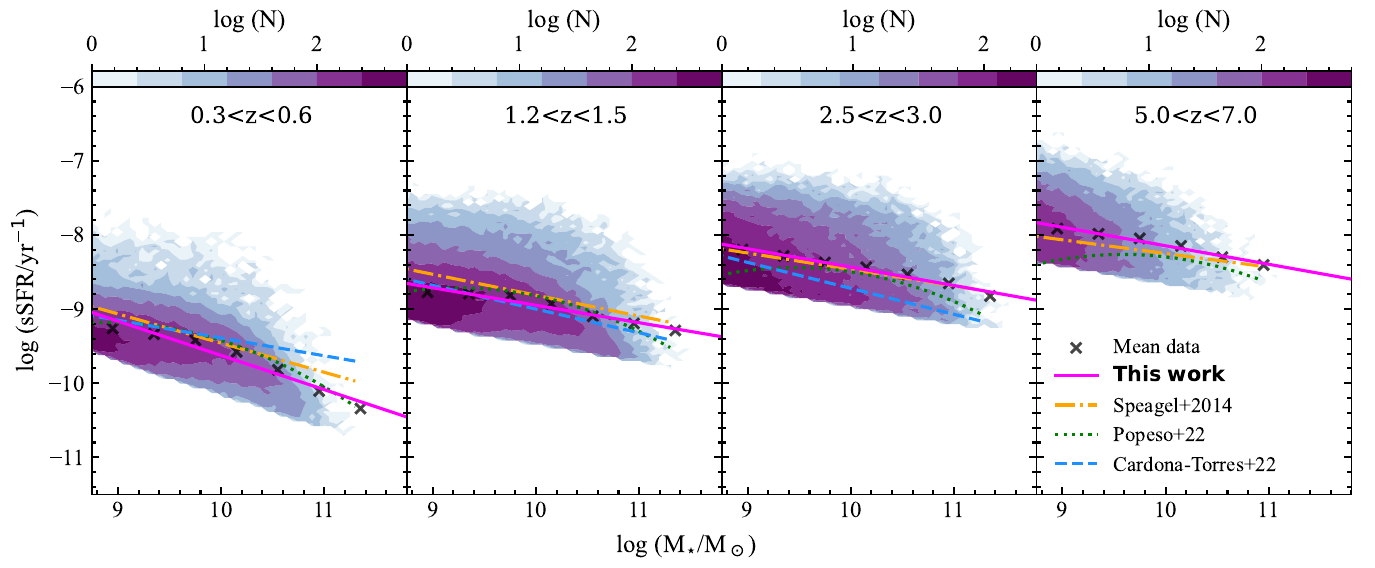}
    \caption{Main sequences of our mock catalogue (magenta solid line) obtained following the selection of star-forming galaxies presented in section \ref{subsec2:sf_selec}. The projected density map represents the selected star-forming galaxies. For comparison, we include the main sequences derived by \citet[orange dashed line]{Speagel+2014}, \citet[green dashed line]{popesso+2022} and \citet[blue dashed line]{cardona-torres+2022}.}
    \label{fig:MS_sim}
\end{figure*}

Figure \ref{fig:MS_sim} shows the sSFR-$M_{\star}$ plane for four different redshift bins, and the corresponding main sequences obtained following the procedure described above. We also include a projected density map (number of galaxies per bins of stellar mass and sSFR) corresponding to the galaxies selected as star-forming. For comparison, we show the main sequences by: \citet{Speagel+2014}, where the authors investigate its evolution in terms of $M_{\star}$, SFR and time (up to $z\sim 6$), using a compilation of 25 studies from the literature; \citet{popesso+2022}, who followed the same procedure as \citet{Speagel+2014} but with a more extensive collection of studies including SFR indicators from the UV, mid-infrared, and far-infrared; \citet{cardona-torres+2022} who used the {\it HST}-CANDELS data to constrain the main sequence in the Extended Groth Strip and characterize a sample of faint SMGs selected at 450 and 850 $\micron$ from the SCUBA-2 Cosmology Legacy Survey. Our predictions are in good agreement with the main sequences by \citet{Speagel+2014} and \citet{popesso+2022}, although the latter one drops at $M_{\star}>10^{11}\;\rm{M}_{\sun}$, a region that is not well constrained by our mock catalogue. Our main sequence is steeper at $z=0.3-0.6$ than that of \citet{cardona-torres+2022}, consistent at $z=1.2-1.5$, and slightly shallower at $z=2.5-3.0$.

\subsection{Infrared Properties of the Galaxies}\label{subsec2:ir_porperties}

In this subsection we describe the methodology followed to establish the infrared emission of the galaxies. Throughout the paper we adopt a \citet{chabrier2003} IMF.

\subsubsection{Dust-obscured SFR and Infrared Luminosities}\label{subsec2:obsc_sfr_ir_luminosity}

To determine the fraction of SFR enshrouded by dust ($f_{\rm obs}={\rm SFR_{IR}/\rm{SFR}}$) for galaxies at $0.5<z<2.5$, we adopt the empirical relation by \citet{whitaker+2017}, who derived $f_{\rm obs}$ for a mass complete sample of star-forming galaxies with $M_{\star} \sim $ $10^{8.7}$ - $10^{11}$\,M$_{\sun}$, selected from the 3D-HST treasury program and \textit{Spitzer}/MIPS $24\;\micron$ maps in the CANDELS fields, finding a strong dependence between $f_{\rm obs}$ and $M_{\star}$ and a negligible evolution with $z$. For $z>2.5$ galaxies we use the results from \citet{dunlop+2017}, where the authors combined ALMA 1.3\;mm observations of the \textit{Hubble} Ultra Deep Field and \textit{HST} data in the UV to trace the IR and total SFRD as a function of $z$. We make a linear fit to their SFRD and SFRD$_{\rm{IR}}$ and estimate $f_{\rm obs}$ as the ratio between both fits (SFRD$_{\rm{IR}}/\rm{SFRD}$). This is later scaled to match the $f_{\rm obs}$ predictions from \citet{whitaker+2017} at $z=2.5$ for the different stellar masses. Finally, for galaxies at $z\leq0.5$ we extrapolate the relation of \citet{whitaker+2017}, with an upper limit of $f_{\rm obs} = 0.75$ for $M_{\star} > 10^{10}$\,M$_{\sun}$ galaxies, motivated by the analytical approximation from \citet{R-P+2020}. The resulting recipe to estimate $f_{\rm obs}$ is:

\begin{align}
\label{ec:fobs1}
&f_{\text{obs}}(M_{\star}>10^{10}\;\text{M}_{\odot}, z\leq0.5) = 0.7\\
\label{ec:fobs2}
&f_{\text{obs}}(M_{\star}<10^{10}\;\text{M}_{\odot}, z\leq0.5) = \frac{1}{1+a\,\text{e}^{b\log(M_{\star}/M_{\odot})}}\\
\label{ec:fobs3}
&f_{\text{obs}}(M_{\star},0.5<z\leq2.5) = \frac{1}{1+a\,\text{e}^{b\log(M_{\star}/M_{\odot})}}\\
\label{ec:fobs4}
&f_{\text{obs}}(M_{\star},2.5<z) = \frac{10^{-0.05z-0.03}}{0.67 [1+a\,\text{e}^{b\log(M_{\star}/M_{\odot})}]},
\end{align}

\noindent where $a$ = (1.96 $\pm$ 0.14)$\times 10^{9}$ and $b$ = -2.277 $\pm$ 0.007.

We use $f_{\rm obs}$ from equations (\ref{ec:fobs1})-(\ref{ec:fobs4}) to estimate SFR$_{\rm{IR}}$ ($=f_{\rm obs}\cdot{\rm{SFR}}$) and derive infrared luminosities for our galaxies following the  relation of \citet{kennicutt_1998} scaled for a \citet{chabrier2003} IMF:

\begin{equation}
    \rm{SFR}_{\rm{IR}} [\rm M_{\sun} yr^{-1}]= 1.09 \times 10^{-10}\, L_{\rm IR}\rm [L_{\sun}].
\label{ec:lir}
\end{equation}

We impose a 2000 M$_{\sun}$ yr$^{-1}$ upper limit to the obscured SFRs. While in large area surveys there are galaxies with apparent flux densities that suggest SFRs larger than this value, follow-up observations have confirmed they are gravitationally amplified systems \citep[e.g.,][]{zavala+2018, vieira+2010, negrello+2010, vieira+2013, harrington+2017}. The intrinsically brightest galaxies found in these surveys have equivalent SFRs close to the limit we adopt \citep[e.g.,][]{karim+2013, barger+2014}.

Table \ref{tab:lightcone} summarizes the general properties of our mock redshift survey.

\begin{table}
\caption{Main properties and range of parameters of the galaxies in our mock redshift survey.}
\centering
{
\begin{tabular}{@{}lcccc@{}} 
\\ \midrule
Area                              & \multicolumn{1}{c}{5.3 deg$^{2}$} \\
\midrule
Parameter                                  & Min.                  & Max.                  \\ \midrule
$z$                          & 0.02                     & 7.0  \\
log($M_{\rm{halo}}${/}M$_{\sun}$) & 9.7     & 13.90   \\
log($M_{\star}${/}M$_{\sun}$)     & 8.75   & 11.83  \\
SFR{[}M$_{\sun}$ yr$^{-1}${]}          & 0.077   & 7847.82      \\ 
SFR$_{\rm IR}${[}M$_{\sun}$ yr$^{-1}${]}          & 0.147   & 2000.0      \\ 
SFR$_{\rm UV}${[}M$_{\sun}$ yr$^{-1}${]}          & 0.062   & 5847.82      \\ \bottomrule
\end{tabular}
}
\label{tab:lightcone}
\end{table}

\subsubsection{Dust Temperature}\label{subsec2:dust_temp}
        
Once $L_{\rm{IR}}$ has been estimated, the corresponding dust temperature ($T_{\rm dust}$) of the galaxies can be inferred. \citet{casey+2018a} used a compilation of dusty star-forming galaxies to model the relation between the intrinsic IR luminosity of galaxies and the observed peak of their SED:

\begin{equation}
    \left\langle\lambda_{\rm peak}(L_{\rm IR})\right\rangle=\lambda_0\left(\frac {L_{\rm IR}}{L_{\rm t}}\right)^\eta,
\label{ec:lpeak_lir}
\end{equation}

\noindent where $\lambda_{0}=102.8 \pm 0.4\; \rm{\micron}$, $L_{\rm t} = 10^{12}\rm\; L_{\sun}$ and $\eta=-0.068 \pm 0.001$. We follow the above relation to associate the intrinsic $L_{\rm{IR}}$ of our galaxies with their corresponding $\lambda_{\rm{peak}}$ and, finally, estimate their $T'_{\rm dust}$ using an approximation to Wien's displacement law:

\begin{equation}
    \lambda_{\rm peak}\approx\frac{\;2.898\times10^3\;\rm  \micron\,\rm {K}}{(T'_{\rm{dust}})^{0.9}}.
    \label{ec:lpeak}
\end{equation}

The $T'_{\rm dust}$ obtained with equation (\ref{ec:lpeak}) must be corrected for heating effects due to the Cosmic Microwave Background (CMB). The temperature of the CMB at $z>4$ approaches and can exceed the temperature of the dust grains at \textit{z} = 0 in the galaxies, rising their temperature. This increase in $T'_{\rm dust}$ boosts the dust continuum emission which implies that the flux densities of galaxies will be higher than without considering this effect. Following \citet{dacunha+2013} the correction in the temperature of the galaxies is given by:                                                                  
\begin{equation}
    T_{\rm{dust}}(z)=({T}{'}_{\rm{dust}}^{4+\beta_{\rm E}}+T_{\rm CMB}^{4+\beta_{\rm E}}\lbrack\; (1+z)^{4+\beta_{\rm E}}-1\; \rbrack)^\frac1{4+\beta_{\rm E}},
    \label{ec:td}
\end{equation}

\noindent where ${T}'_{\rm{dust}}$ is the dust temperature of the galaxy in absence of the CMB, $T_{\rm{CMB}}$ = 2.725 K is the CMB temperature at $z=0$ and $\beta_{E}=1.8$ is the dust spectral emissivity index \citep[value adopted from][]{casey+2018a}.

    \subsubsection{Observed Flux Density}\label{subsec2:obs_flux}

To estimate the observed flux density ($S_{\nu}$) of our mock galaxies we assume that their spectral energy distribution (SED) is well modeled by a modified black body given by: 
        
\begin{equation}
    f(\nu,T)\propto(1-{\rm e}^{-\tau(\nu)})B_\nu(T),
    \label{ec:gb}
\end{equation}

\noindent where $\tau(\nu)=(\nu / \nu_{0})^{\beta_{\rm{E}}}$ is the optical depth, $\nu_{0}$ = 3 THz is the frequency at which the modified black body changes from optically thick to optically thin and $B_{\nu}(T)$ is the Planck function. The observed flux density of our galaxies at frequency $\nu$ is estimated through the relation:

\begin{equation}
    S_\nu=\frac{1+z}{4\pi D_{\rm{L}}^{2}}L_{\rm{IR}}\frac{f_{\nu(1+z)}}{\int f_{\nu'}\;d\nu'},
    \label{ec:kcorr}
\end{equation}

\noindent where $f_{\nu}$ is the intrinsic SED of galaxies (equation \ref{ec:gb}), $L_{\rm{IR}}$ is their infrared luminosity, $z$ is their redshift and $D_{\rm{L}}$ is the luminosity distance. The total infrared luminosity is defined as the integral of the SED from 8 to 1000 $\micron$.

\subsection{Gravitational Lensing}\label{subsec2:grav_lens}

To introduce gravitational lensing effects in our mock redshift survey, we follow \citet{narayan+1995} and consider amplifications produced by dark-matter haloes in the line of sight (under a point-mass model approximation) on the background galaxies.

In order to identify halo-galaxy alignments, we set a search radius $r_{\rm{b}}=5$ arcsec, with origin at the position of each halo. We find that background sources at angular distances larger than this value will have amplifications smaller than 20 per cent for the most massive haloes ($M_{\rm{halo}} > 10^{14}$\,M$_{\sun}$), and therefore do not have a strong impact in our results. For each halo with identified background galaxies, we can calculate its corresponding Einstein radius assuming a point-mass approximation \citep{narayan+1995}:

\begin{equation}
  \theta _{\rm{E}}=\left (\frac{4GM}{c^{2}} \frac{D_{\rm{ds}}}{D_{\rm{d}}D_{\rm{s}}} \right )^{1/2},
  \label{ec:ein_rad}
\end{equation}

\noindent with $G$ the gravitational constant, $c$ the speed of the light, and $D_{\rm{d}}$, $D_{\rm{ds}}$ and $D_{\rm{s}}$ the angular distances between observer and lens, lens and source, and observer and source, respectively. The angular separation ($\theta$) of the lensed images is given by:

\begin{equation}
\theta_{\pm}=\frac{1}{2}\left(\beta_{\rm{l}} \pm \sqrt{\beta_{\rm{l}}^{2}+4\theta_{\rm{E}}^{2}}\right),
  \label{ec:angle_im}
\end{equation}

\noindent with $\beta_{\rm{l}}$ being the angular separation between the optical axis between the observer and the lens, and the true source position. Combining equations (\ref{ec:ein_rad}) and (\ref{ec:angle_im}), the magnification of the image is:

\begin{equation}
\mu_{\pm}=\left [1-\left (\frac{\theta_{\rm{E}}}{{\theta_{\pm }}} \right )^{4} \right]^{-1}  ,
  \label{ec:magnification}
\end{equation}

\noindent with the total magnification given by $\mu=\left|\mu_{+}\right|+\left|\mu_{-}\right|$.

 %----------------SECTION 3--------------------------
 
\section{Tests of the Mock Catalogue}\label{sec3:sim_characterization}
    
In order to characterize the performance of our mock redshift survey, we measure the predicted cumulative number counts, star-formation rate density, total infrared luminosity function and redshift distribution, and compare them with observational results. 

\subsection{Number Counts}
\label{subsec3:ncounts}

\begin{figure}
  \centering
\includegraphics[width=\columnwidth]{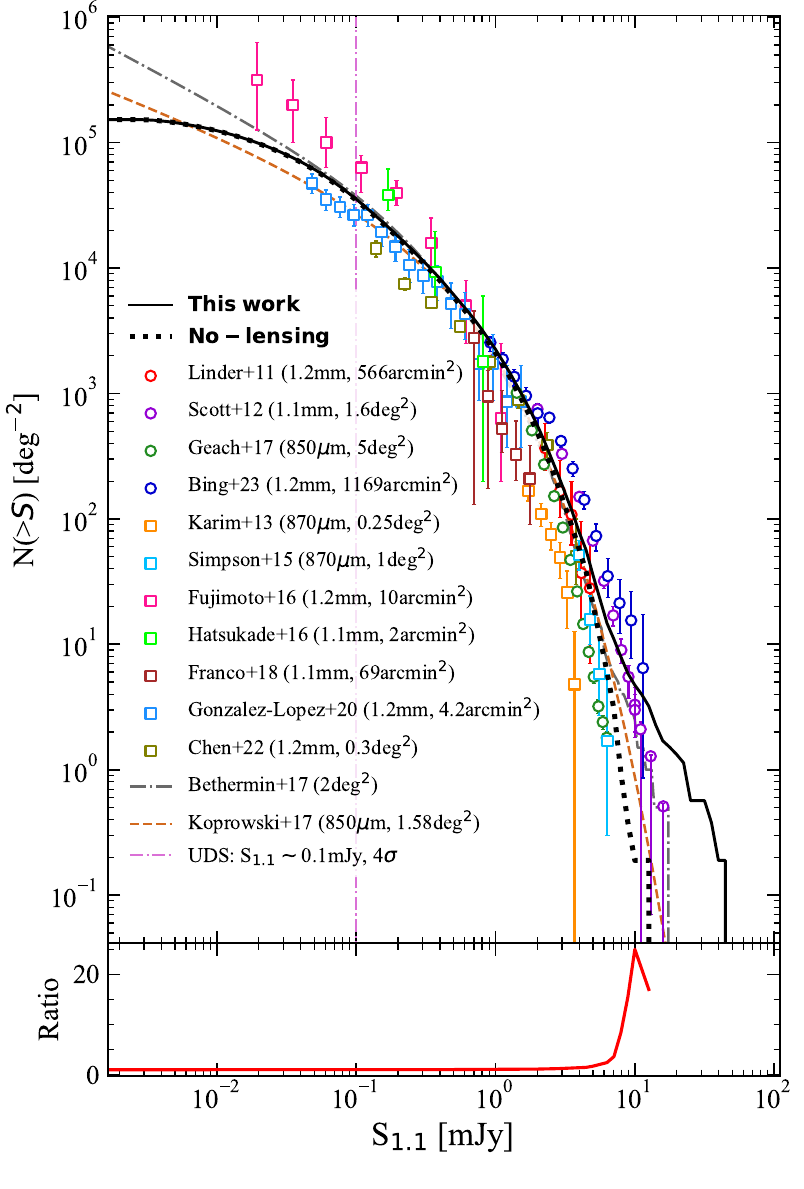}
    \caption{Top panel: Integrated number counts at 1.1 mm from our mock redshift survey before (black dotted line) and after (black solid line) introducing the effects of gravitational lensing. We compare our results with different interferometric and single-dish observations (squares and circles, respectively). We also represent the number counts obtained by integrating the luminosity function of \citet[brown dashed line]{koprowski+2017} and those from the 2 deg$^{2}$ SIDES simulation \citep[][gray dashed-dotted line]{bethermin+2017}. The vertical line indicates the expected 4$\sigma$ detection limit for the UDS/TolTEC survey. Bottom panel: ratio between the number counts measured in our mock redshift survey with and without gravitational lensing effects.}
    \label{fig:ncounts1.1}
\end{figure}

Figure \ref{fig:ncounts1.1} shows the number counts estimated from our 5.3 deg$^{2}$ mock redshift survey, both with and without the inclusion of gravitational lensing effects. Our results are compared with different interferometric and single-dish observations \citep{Lindner+2011, scott+2012, karim+2013, simpson+2015, fujimoto+2016, hatsukade+2016, geach+2017, gonzalez-lopez+2020, franco+2020, chen+2022, bing+2023}, extrapolating the observed number counts from other wavelengths to 1.1 mm assuming $\beta_{\rm E} = 1.8$ as in section \ref{subsec2:dust_temp}. Also included are the number counts measured from the SIDES numerical simulation \citep{bethermin+2017}, and those predicted by integrating the IR luminosity function from \citet{koprowski+2017}, which is based on the SCUBA-2 observations of the UDS and COSMOS fields (from the larger SCUBA-2 Cosmology Legacy Survey, \citealt{geach+2017}) and the deep ($1\sigma_{\rm 1.3} \approx 35\,\mu$Jy) ALMA observations of the 4.5 arcmin$^2$ {\it Hubble Ultra Deep Field} \citep[HUDF,][]{dunlop+2017}.

Given the current uncertainties and the dispersion between the different observational results, our mock catalogue accurately reproduces the measured number counts. At the faint end ($S_{\rm 1.1 }\lesssim 0.1$\, mJy), our predictions fall below the 1.2 mm ALMA results from \citet{fujimoto+2016}. Their measurements, however, were derived from 120 pointings (mainly archival data) towards different astronomical sources, potentially biasing their results. Furthermore, the surveyed area available to estimate the number counts at $S_{\rm 1.1 }\lesssim 0.1$\, mJy was $\sim$ 2 sq.arcmin, which may introduce sampling variance uncertainties. On the other hand, at $S_{\rm 1.1 }\sim 0.1$\, mJy, the 1.2 mm ALMA number counts from \citet{gonzalez-lopez+2020} and \citet{chen+2022} fall slightly below our predictions. \citet{gonzalez-lopez+2020} observed 4.2 arcmin$^{2}$ of the Hubble Ultra Deep Field, and \citet{chen+2022} identified faint sources around ALMA calibrator maps (ALMACAL project). Both studies are based on the analysis of small-area individual maps, and are also potentially affected by sampling variance effects\footnote{We note that, although the total integrated area of the ALMACAL project is 0.3 deg$^{2}$, the deeper $S_{\rm 1.2 } < 0.5$\, mJy area is $\lesssim$ 80 sq.arcmin with $\sim 1$ sq.arcmin individual maps.}.

The bright end of the number counts has been mainly traced by single-dish observations, e.g. the 5 sq.degree SCUBA-2 Cosmology Legacy Survey including seven independent fields \citep[0.07 - 2.22 sq.deg; ][]{geach+2017}, the 1.6 sq.degree AzTEC JCMT-ASTE survey of seven blank-fields (0.08 - 0.72 sq.deg) and 3.1 sq.degree towards 37 potentially overdense regions \citep[][]{scott+2012}, and the recent 0.32 sq. degree NIKA2 Cosmological Legacy Survey including GOODS-N and COSMOS. Interferometric observations have explored this bright end through follow-up observations of the brightest sources detected by single-dish telescopes. For example, \citet{simpson+2015} presented $870 \micron$ ALMA pointed observations of 30 galaxies selected at $850 \mu$m from the SCUBA-2 Cosmology Legacy Survey, finding that 61 per cent of the selected sources were the result of blending two or more fainter DSFGs. \citet{karim+2013} reached a similar conclusion through $870 \micron$ ALMA follow-up observations of 122 SMGs selected from the Large Apex BOlometer CAmera (LABOCA) Extended \textit{Chandra} Deep Field South submillimetre survey (LESS), with all of the brightest sources ($S_{\rm 870} \gtrsim 12$\, mJy) comprising emission from multiple fainter galaxies. At the bright end ($S_{1.1}\gtrsim$\,5 mJy), our mock catalogue starts to deviate from the general trend of the observed number counts and from our predictions if no gravitational lensing effects are introduced. This transition therefore indicates where gravitational lensing effects starts to have a strong impact on the measured number counts, although a small contribution from intrinsically bright galaxies is also expected given the larger area of our mock redshift survey compared to the observational measurements \citep[i.e., observing larger areas increases the probability of finding brighter sources and strongly lensed galaxies;][]{clements+2010}. The bottom panel of Figure \ref{fig:ncounts1.1} shows the ratio between our mock redshift survey with and without lensing: at $S_{1.1}>5$ mJy, the number counts start to be dominated by gravitationally lensed galaxies, which can increase the un-lensed number counts by a factor of up to 26 at $S_{1.1}\sim10$ mJy. 

The number counts from our mock redshift survey are also in good agreement with those from SIDES \citep{bethermin+2017} at $S_{\rm 1.1} \gtrsim 0.05$\, mJy. The difference at lower flux densities can be the result of the different mass resolution between the mock catalogues: SIDES includes galaxies with $M_{\star} \geq 10^{8}$ M$_{\sun}$ whereas our new redshift survey has a resolution of $M_{\star} \geq 10^{8.75}$ M$_{\sun}$. It is worth noting that while the Bolshoi-\textit{Planck} simulation resolves haloes down to $V_\text{peak}\sim 60$ km/s, which corresponds to haloes of $M_\star \sim 4\times 10^{7}$ M$_{\sun}$, the scatter (the product of intrinsic evolution plus random errors) around the $M_{\star}-V_\text{peak}$ relation allows for a conservative completeness in stellar mass terms down to $M_{\star} \geq 10^{8.75}$ M$_{\sun}$. 

\begin{figure*}
\makebox[\textwidth]{\includegraphics[width=\textwidth]{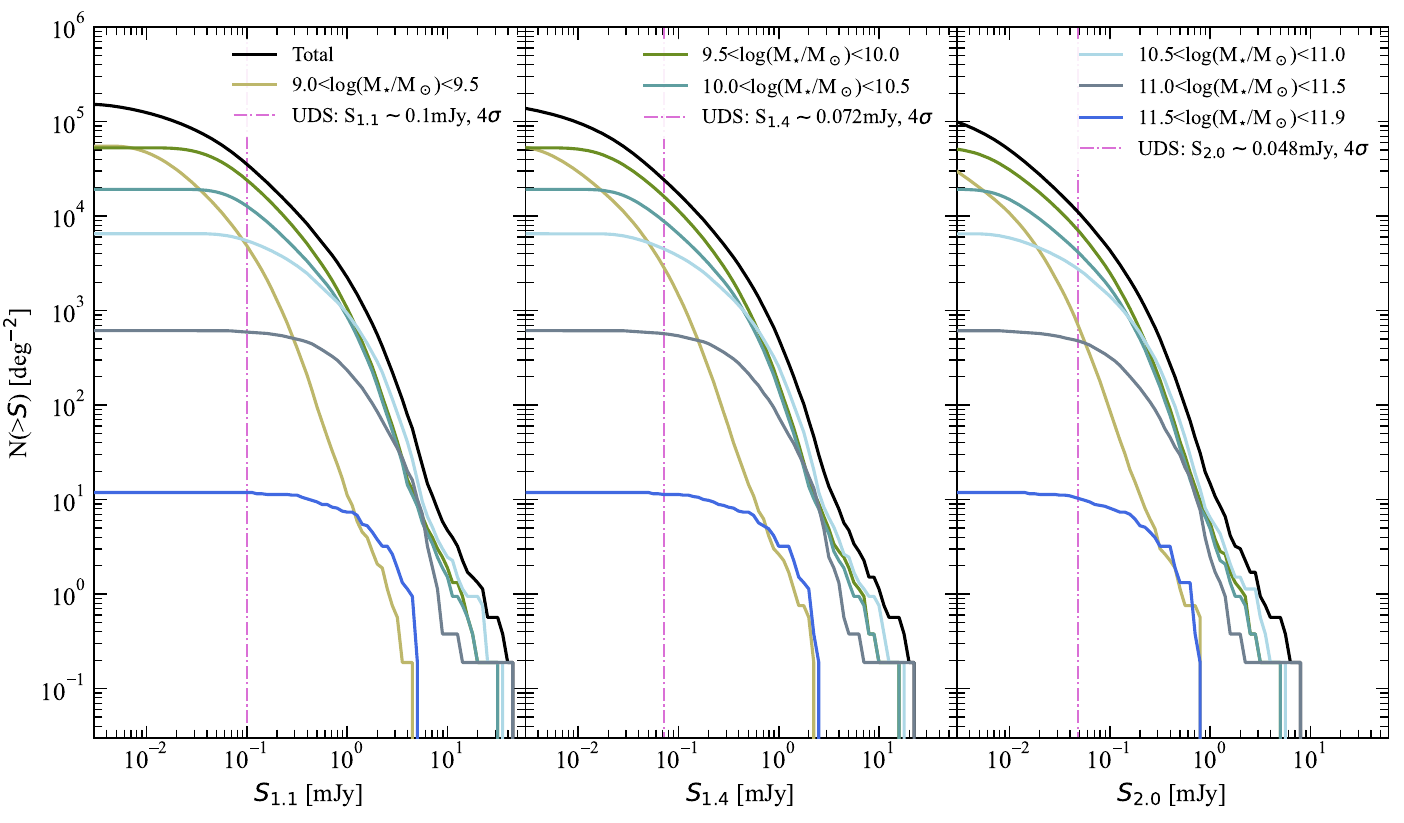}}
\caption{Number counts at 1.1, 1.4, and 2.0 mm. The black solid line represents our measurement accounting for all the galaxies in the mock redshift survey. The color solid lines show the contributions to the number counts where we separate the galaxies by stellar mass ranges. The vertical lines are the 4$\sigma$ detection limits in the UDS surveys.}
    \label{fig:ncounts_lambda_sm}
\end{figure*}

In order to explore the relative contribution of galaxies with different stellar masses to the overall population of DSFGs, Figure \ref{fig:ncounts_lambda_sm} shows the 1.1, 1.4, and 2.0 mm number counts measured from our mock redshift survey for different $M_{\star}$ bins. As expected, a general trend is seen for the three wavelengths, with the faint end (e.g. $S_{1.1} \lesssim 0.1$\, mJy) being dominated by the lower mass log($M_{\star}$/M$_{\sun}) = 9.5-10.0$ population, and more massive galaxies increasingly contributing towards larger flux densities. Interestingly, the interval $S_{1.1} \sim 1 - 10$\, mJy, which has been mainly probed by typical single-dish surveys, has an almost equivalent contribution by the three stellar-mass bins in the range log($M_{\star}$/M$_{\sun}) = 9.5-11.0$, with a small contribution from log($M_{\star}$/M$_{\sun}) = 11.0-11.5$ galaxies and the most massive log($M_{\star}$/M$_{\sun}) > 11.5$ population having a negligible contribution to the number counts at all flux densities. Finally, Figure \ref{fig:ncounts_lambda_sm} indicates that the $S_{1.1} \sim 0.1$\, mJy detection limit of the TolTEC Ultra Deep Legacy Survey will allow us to probe the relatively low mass ($9.5\lesssim{\rm log}(M_{\star}/{\rm M}_{\sun}) \lesssim 10$) population of DSFG, and study their properties in a sample of a few tens of thousands of galaxies.

\subsection{SFR History}

\begin{figure}
\includegraphics[width=\columnwidth]{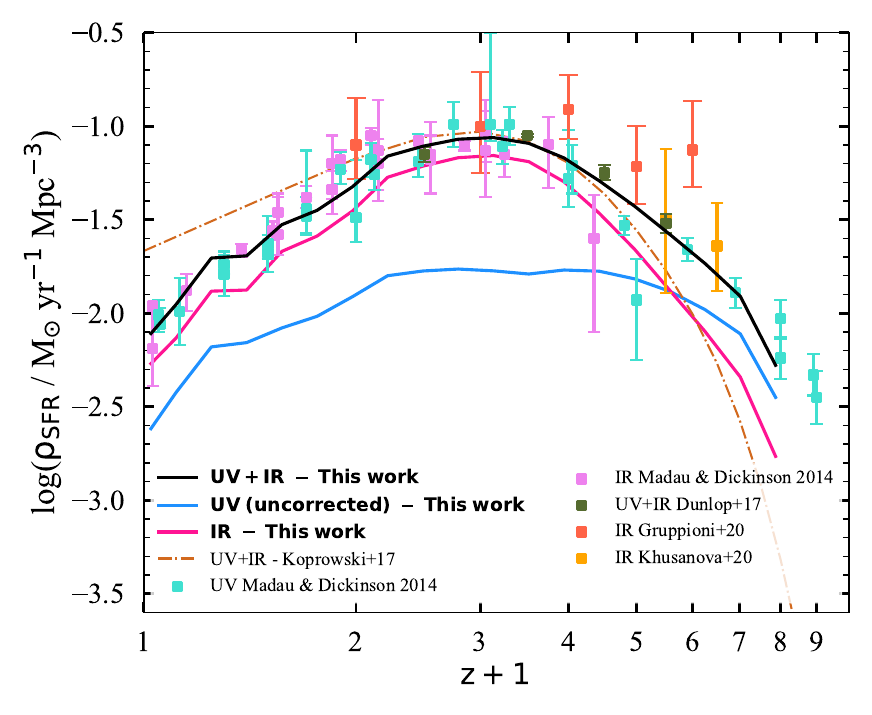}
    \caption{Total SFRD measured in our mock redshift survey (black solid line), and its relative contribution from the unobscured UV and dust-obscured IR components (blue and pink solid lines respectively). Our results are in good agreement with the observational data from \citet[pink and cyan squares]{madau_dickinson2014}, \citet[olive green squares]{dunlop+2017}, \citet[yellow squares]{khusanova+2020} and \citet[orange squares]{gruppioni+2020}. We note that the compilation of UV measurements from \citet{madau_dickinson2014} have been corrected for dust attenuation in order to recover the total SFRD. We also plot the SFRD estimate by \citet[brown dashed-dotted line]{koprowski+2017}.}
    \label{fig:sfrd}
\end{figure}

The total SFRD (IR+UV) as a function of $z$ measured from our mock redshift survey is presented in Figure \ref{fig:sfrd}, showing the relative contributions from the IR and UV (including the effects of dust attenuation) SFR. Our predictions are consistence with different observational results and models from the literature \citep[e.g.,][]{zavala+2021}, in particular with the fact that the dust-obscured star formation activity dominates the star formation history of the Universe for the last $\sim$ 12.4 Gyr (i.e. since $z \approx 4.5$). For comparison, Figure \ref{fig:sfrd} includes: the compilation of UV and IR measurements from \citet{madau_dickinson2014}, where the UV observations have been corrected for dust attenuation; the total SFRD presented in \citet{dunlop+2017}, derived combining deep ALMA observations and rest-frame UV data of the 4.5 arcmin$^{2}$ HUDF; the results from \citet{khusanova+2020}, who estimated the SFRD at $z = 4.5$ and 5.5 from the UV-selected galaxies in the ALMA Large Program to INvestigate CII at Early Times (ALPINE); and the \citet{gruppioni+2020} SFRD derived from serendipitous detections around the ALPINE targets. Also shown is the total SFRD from \citet{koprowski+2017}, estimated by integrating their measured IR luminosity function and including data from \citet{parsa+2016} to incorporate the UV contribution.

\subsection{Redshift Distribution}
\label{subsec3:zdist}
\begin{figure}
  \centering
    \subfloat{\includegraphics[width=\columnwidth]{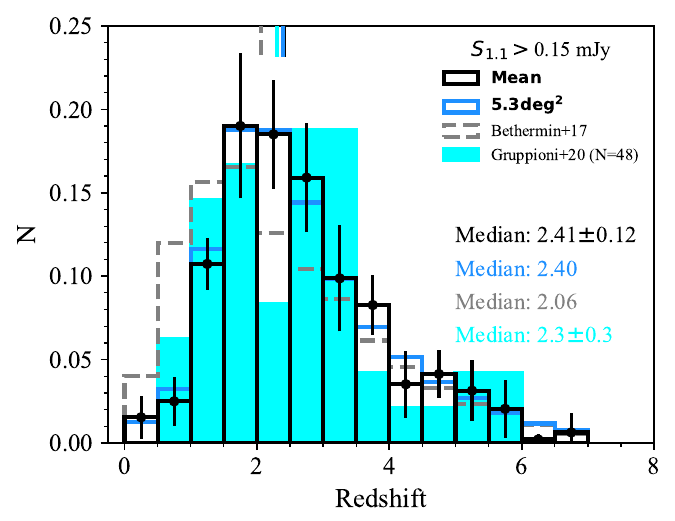}}

    \subfloat{\includegraphics[width=\columnwidth]{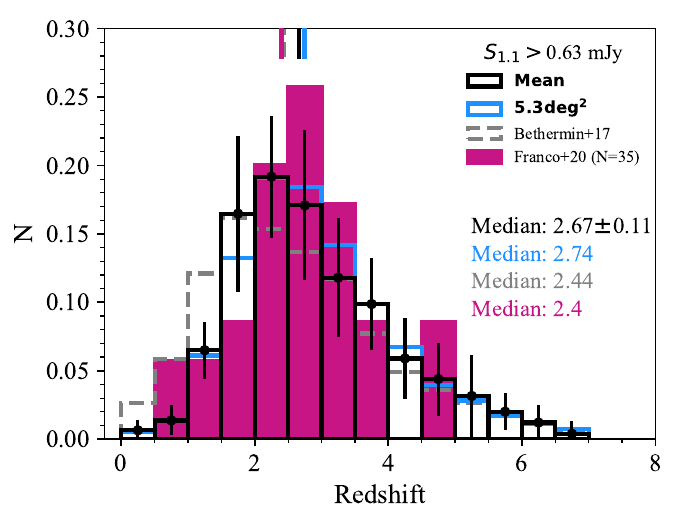}}

    \subfloat{\includegraphics[width=\columnwidth]{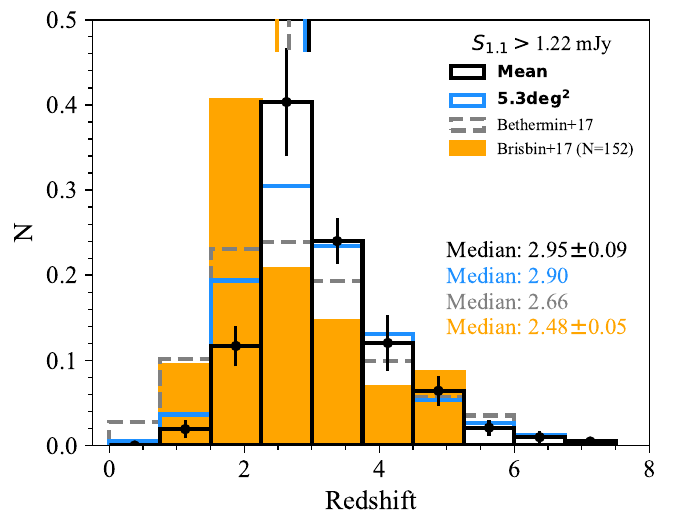}}
  \caption{Redshift distributions measured in \citet[][$S_{1.1}>0.15$ mJy, cyan histogram]{gruppioni+2020}, \citet[][$S_{1.1}>0.63$ mJy, magenta histogram]{franco+2020} and \citet[][$S_{1.1}>1.22$ mJy, yellow histogram]{brisbin+2017}, compared to those predicted by our 5.3 sq. degree mock redshift survey (blue histogram) and the average of 10 mock surveys matching the observed areas and depths (black histogram). The black error bars in each redshift bin indicate the 1$\sigma$ dispersion from the 10 realizations. The gray dashed histogram shows the redshift distributions predicted from the SIDES simulation \citep{bethermin+2017}. The short vertical lines at the top of the plots indicate the median values for each distribution.}
  \label{fig:zdist}
\end{figure}

To test the redshift distribution of DSFGs predicted by our mock catalogue, we reproduce, in terms of depth and total area, three recent millimeter wavelength surveys that probe the faint and bright population of DSFG: \citet{brisbin+2017}, \citet{gruppioni+2020}, and \citet{franco+2020}. For each of these surveys, we generate ten independent realizations and estimate the average redshift distribution. Figure \ref{fig:zdist} presents our results and the comparison is discussed below. As an additional reference, the redshift distribution measured on the complete 5.3 sq. degree mock redshift survey and using the total area of SIDES are also shown.

As it has been mentioned, \citet{gruppioni+2020} studied a sample of 56 serendipitously detected sources in the ALPINE survey of high redshift ($z=4.4-5.9$) galaxies. ALPINE reached a sensitivity equivalent to that expected for the TolTEC UDS survey ($S_{\rm 1.1} \sim 0.1$ mJy), and the total area where serendipitous sources were detected (i.e. excluding the central region of the high-$z$ target) corresponds to 25 sq. arcmin. \citet{gruppioni+2020} estimated a median redshift of $z_{\rm{med}}=2.3\pm0.3$ for their sample, and our mock redshift surveys predict $z_{\rm{med}}=2.41\pm0.12$ (a similar value is measured over the whole 5.3 sq.deg mock redshift survey). While these median redshifts are in good agreement, it should be noted that the results from \citet{gruppioni+2020} may be biased by groups of galaxies gravitationally lensing the ALPINE targets or sources associated to these high-$z$ systems. These potential biases, combined with sampling variance effects due to the small survey area, can impact the overall shape of the observed redshift distribution. On the other hand, the SIDES simulation predicts a larger number of low-$z$ DSFGs resulting in a lower median redshift of $z_{\rm{med}}=2.1$.

In the case of \citet{franco+2020}, the authors took advantage of 3.6 and 4.5 $\mu$m {\it Spitzer}/IRAC and 3 GHz Very Large Array data to extend the catalogue of sources from the 69 sq. arcmin GOODS-ALMA survey \citep{franco+2018}, reducing the detection limit down to $S_{\rm 1.1} \geq 0.63$\, mJy. The final catalogue includes 35 DSFGs with photometric and spectroscopic redshifts in the range $z = 0.65-4.73$. Interestingly, including the fainter population, identified by means of the near-IR and radio data, decreased the median redshift of the sample from $z_{\rm{med}}=2.73$ to $z_{\rm{med}}=2.40$. In our case, the mean distribution of our mock redshift surveys has a median value of $z_{\rm{med}}=2.67\pm0.11$, and $z_{\rm{med}}=2.74$ for the complete mock redshift survey. Our predictions are therefore slightly higher but still consistent with the observational results, particularly with the original catalogue of $S_{\rm 1.1} \geq 0.88$\, mJy ALMA sources. The SIDES mock catalogue, on the other hand, shows a larger population of lower redshift galaxies and predicts $z_{\rm{med}}=2.44$, in better agreement with the extended catalogue of \citet{franco+2020}. As in the case of \citet{gruppioni+2020}, the redshift distribution shows sensitivity to variations around the peak value due to the small map sizes, where the cosmic variance dominates.

Finally, \citet{brisbin+2017} presented ALMA 1.25 mm follow-up observations of 129 sources detected with $S_{\rm 1.1} \gtrsim 3.5$\, mJy in the 0.72 sq. degree AzTEC/ASTE maps of COSMOS (FWHM $\sim 30\,$ arcsecond). The high sensitivity ($1\sigma_{\rm 1.25} \sim 0.15\,$ mJy) and angular resolution of the ALMA observations resulted in the detection of 152 galaxies, 143 of them with either spectroscopic or accurate photometric redshifts, and with a median redshift $z_{\rm{med}}=2.48\pm0.05$. Our mock redshift survey predicts a considerably higher value of $z_{\rm{med}}=2.95\pm0.09$ while the SIDES mock catalogue predicts a closer value of $z_{\rm{med}}=2.66$.

\subsection{Infrared luminosity function}
    
\begin{figure*}
\includegraphics[width=\textwidth]{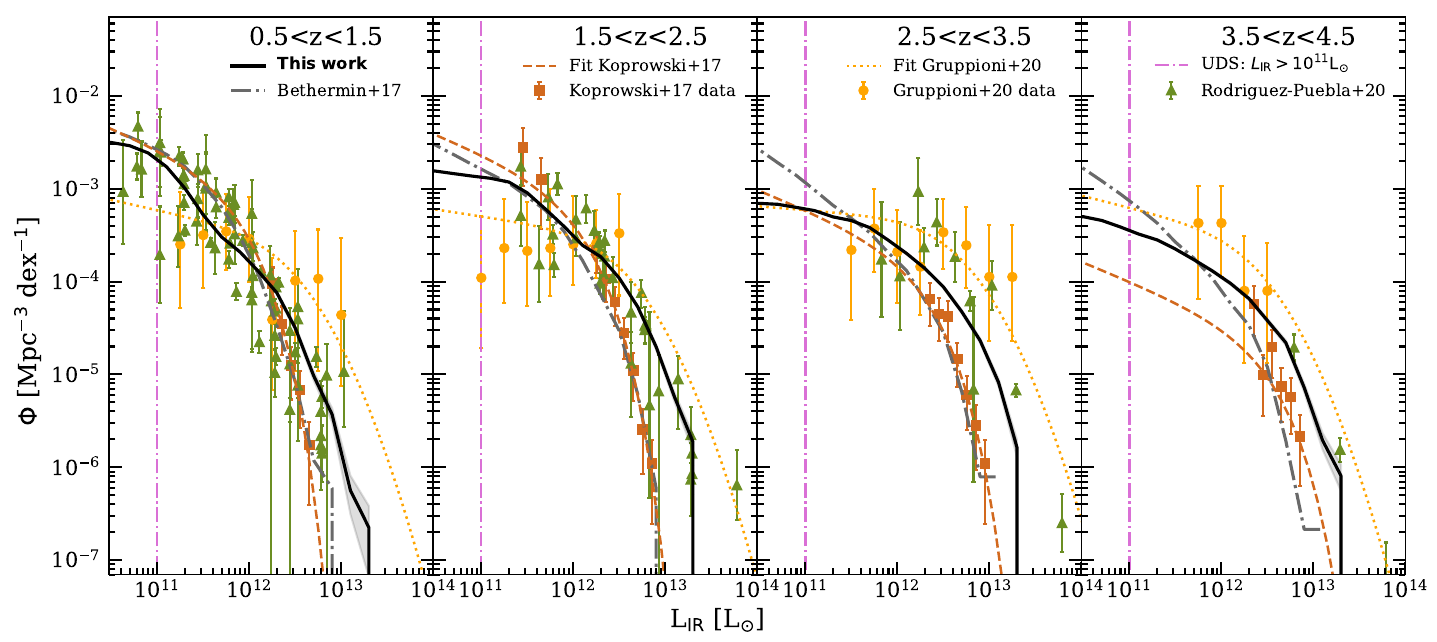}
    \caption{Infrared luminosity function measured in our 5.3 deg$^{2}$ mock redshift survey (black solid line, with shaded regions indicating the $1\sigma$ Poisson error). The brown squares represent the sample by \citet{koprowski+2017} and brown dashed line they fit. The yellow circles are the data by \citet{gruppioni+2020} and the yellow dotted line their fit. The green triangles are the compilation data by \citet{R-P+2020}  calibrated to the same cosmology. The \citet{bethermin+2017} simulation is indicated with the gray dashed-dotted line. The vertical line indicates the limit in luminosity that the UDS/TolTEC survey will reach at S/N=4.}
    \label{fig:lf}
\end{figure*}

To measure the infrared luminosity function in our mock redshift survey we directly count the number of galaxies in $L_{\rm IR}$ and $z$ bins. Figure \ref{fig:lf} shows the results compared to different observational studies. We find that our mock redshift survey predicts more luminous galaxies than the infrared luminosity function from \citet{koprowski+2017}, which is based on the $\sim$1.5 sq. degree SCUBA-2 Cosmology Legacy Survey \citep{geach+2017} and the deep 1.3 mm ALMA observations of the HUDF \citep[covering 4.5 arcmin$^{2}$;][]{dunlop+2017} to constrain the faint end. \citet{gruppioni_pozzi2019}, however, suggest that the single SED template used by \citet{koprowski+2017} to estimate luminosities may have biased their results. Furthermore, the 850 $\mu$m SCUBA-2 observations are less sensitive to `warm' galaxies, a population that tends to dominate the bright end of the luminosity function, as suggested by 70, 100 and 160 $\micron$ {\it Herschel} observations \citep{gruppioni+2013}. On the other hand, compared to our predictions, the luminosity function derived from the analysis of serendipitous detections around the ALPINE targets \citep{gruppioni+2020}, shows a systematic excess of luminous galaxies ($L_{\rm IR} \geq 10^{12} {\rm L}_\odot$) through all redshift bins. Towards fainter luminosities, however, their curve flattens and starts falling below our predictions, particularly at $z<2.5$. Finally, \citet{R-P+2020} used a compilation of several observational results at $z<4.2$ (see Table 2 of that work) to constrain analytical models of the infrared luminosity function. Their compilation includes data from \textit{Spitzer} (24 and $70\micron$), \textit{AKARI} ($90 \micron$), \textit{Herschel} (70, 100, 160, 250, 350 and $500 \micron$), and SCUBA-2 ($450 \micron$). The luminosity function predicted by our mock redshift survey is in very good agreement, given the uncertainties and dispersion of the data, with this multi-wavelength compilation.

To better constrain the evolution of the luminosity function, larger statistical samples are still required, particularly at the higher redshifts (i.e. $z > 2.5$). In \S \ref{sec:toltec_lf} we take advantage of our mock redshift survey to make predictions on the overall improvement that the TolTEC UDS will represent in our understanding of the evolution of the luminosity function.

    \subsection{Clustering characterization}

To characterize the clustering properties of the DSFG in our mock redshift survey we estimate their angular two-point auto-correlation function ($\omega$).
This statistical tool measures the excess probability of finding a pair of galaxies separated by an angular scale $\theta$, compared to a uniform non clustered distribution. To compute $\omega$ we use the numeric estimator presented in \citet{Landy1993}: 

\begin{equation}
    \omega(\theta) =
    \frac{DD(\theta)-2DR(\theta)+RR(\theta)}{RR(\theta)}
\end{equation}

\noindent where $DD$, $RR$ and $DR$ are the number of galaxy pairs in the clustered, uniform and cross-matched distributions respectively. To improve the efficiency of the calculations, considerably reducing computing time, the algorithm of \citet{He2021} was adapted to analytically estimate $RR$. We also include integral constraint corrections \citep{Roche1999}, which are particularly important to consider when $\omega$ is measured at angular scales of order of the map size.

\begin{figure}
  \centering
    \subfloat{\includegraphics[width=0.46\textwidth]{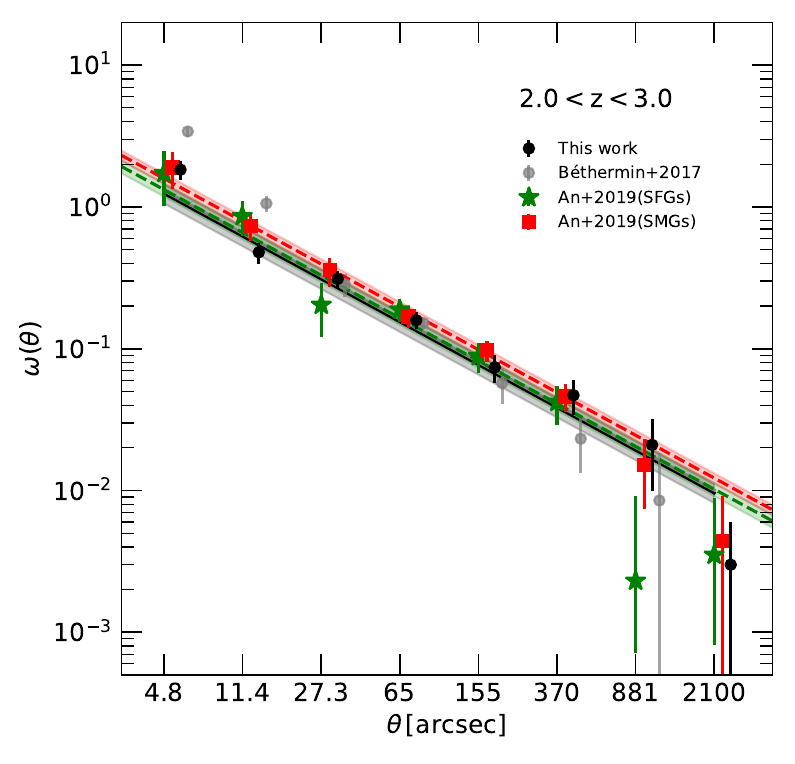}}
    
  \caption{Angular two-point autocorrelation function of sources in a 2 sq. degree area of our mock redshift survey (black dots). A redshift bin $z = 2-3$ and a stellar masses $M_{\star}>10^{10.5}\, \textrm{M}_{\sun}$ are adopted to allow a direct comparison with the observational results from \citet{An2019}, one of the most statistically robust clustering measurements of faint SMGs available to date. Straight lines correspond to power-law fits ($\omega(\theta) = A\theta^{-0.8}$) to the different data sets, with shaded regions indicating $1\sigma$ bootstrapped errors, showing a general good agreement given the current uncertainties. The clustering of DSFGs in the SIDES mock catalogue is also included as a complementary comparison. All measurements correspond to the same angular scales considered in \citet{An2019}, but have been slightly shifted for a better visualization.}
  \label{fig:AnBeth}
\end{figure}

\citet{An2019} studied the clustering properties of main-sequence Star-Forming Galaxies (SFGs) and from $\sim 5800$ SMGs (with $M_{\star}>10^{10.5}\, \textrm{M}_{\sun}$) identified in the 2 sq. degree COSMOS field using a machine-learning technique trained with deep SCUBA-2 and ALMA data. Due to the large mapped area, sample size, and available optical/IR photometric redshifts, this currently represents the most statistical robust clustering measurements of the DSFG population. Figure \ref{fig:AnBeth} compares the angular two-point auto-correlation functions measured by \citet{An2019} in the $z = 2 - 3$ redshift bin, and that of $M_{\star}>10^{10.5}\, \textrm{M}_{\sun}$ sources in a 2 sq. degree area of our mock redshift survey. The uncertainties in our measurements are estimated through a bootstrap analysis of 50 maps drawn from the larger 5.3 sq. degree mock redshift survey, therefore taking into account cosmic variance effects due to the finite size of the survey area. As a complementary comparison, we estimate the clustering of DSFG from the SIDES simulation, using 1 sq. degree maps for the bootstrap analysis due to the smaller size of the simulation.

Figure \ref{fig:AnBeth} shows that the clustering properties of the DSFG in our mock redshift survey are in good agreement with those measured for the SMGs and SFGs identified in the COSMOS field. We further fit a power-law of the form $\omega(\theta) = A\theta^{-0.8}$ to the different data sets, and find equivalent clustering amplitudes (within the uncertainties) between them: $A = 4.38\pm0.67$ for our mock catalogue, and $A = 4.66\pm0.59$ and $5.52\pm0.38$ for the SFGs and SMGs samples in \citet{An2019}. As mentioned above, our error bars are larger since they are estimated from 50 realizations of 2 sq. degree maps and hence consider cosmic variance. A similar agreement is found for the other two redshift bins ($z = 1-2$ and $z = 3-5$) presented in \citet{An2019}, although affected by smaller statistics at the higher redshifts.

%----------------------------SECTION 4--------------

\section{Predictions for the UDS/TolTEC survey} \label{sec3:pred_toltec_areas}

In this section we take advantage of the mock redshift survey developed in \S \ref{sec2:simulation} to make predictions of the constrains that the TolTEC UDS will make on the number counts, infrared luminosity function, and redshift distribution of the DSFG population. The UDS will consist of deep ($1\sigma_{\rm 1.1} \approx 0.025$\, mJy) observations on a larger 0.5 sq. degree area of COSMOS and smaller ($\gtrsim$ 200 arcmin$^{2}$) maps towards other CANDELS fields. We adopt these two map sizes and, for each area, generate 10 independent realizations in order to provide mean values and sampling variance for each observable.

\subsection{UDS/TolTEC number counts} 

\begin{figure*}
\makebox[\textwidth]{\includegraphics[width=\textwidth]{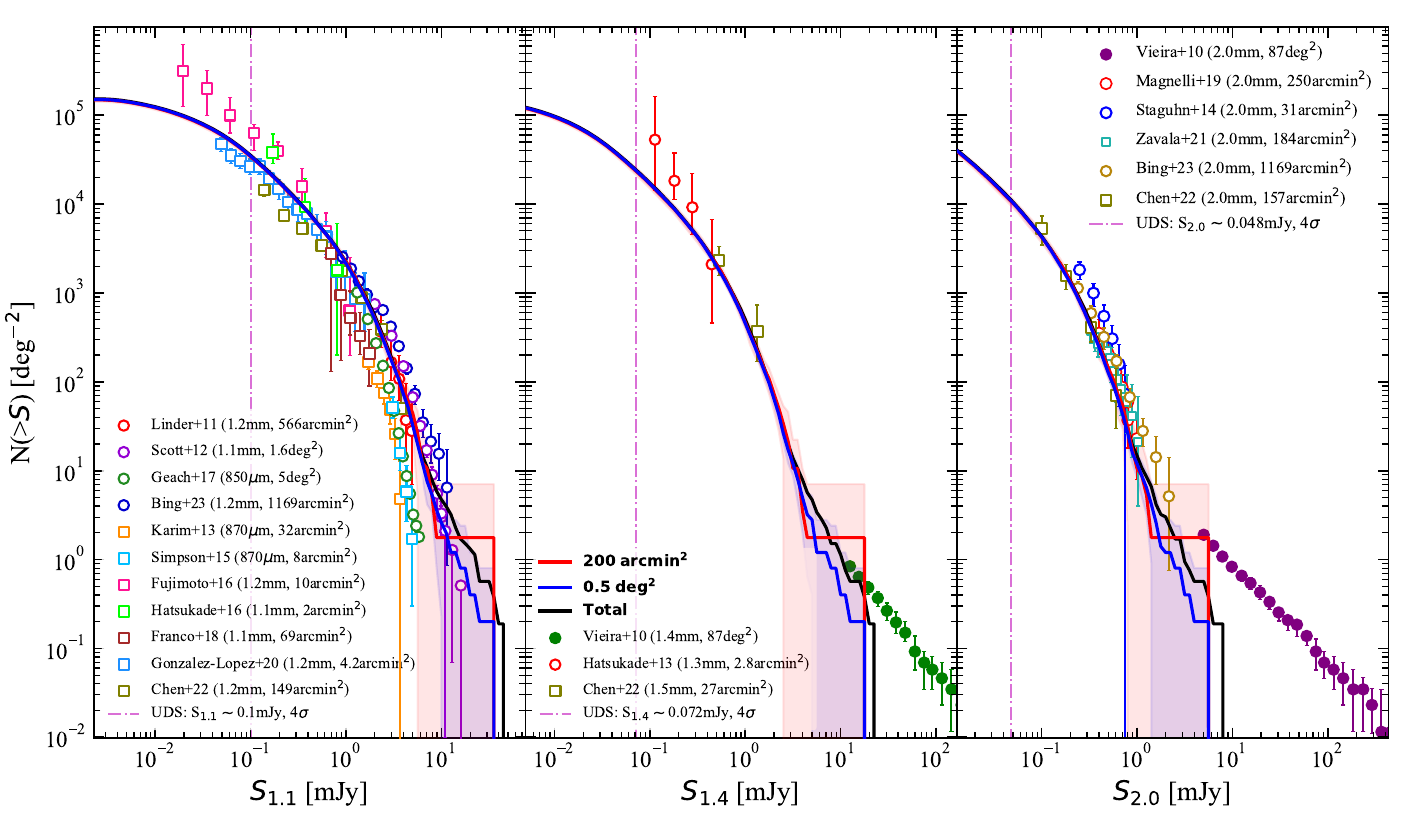}}
\caption{ Mean 1.1, 1.4 and 2.0 mm number counts predicted by our mock redshift survey for the areas of the UDS/TolTEC survey, i.e. 200 arcmin$^{2}$ (red) and 0.5 deg$^{2}$ (blue). The shaded area represent the $1\sigma$ dispersion from ten realizations, and the vertical lines indicate the expected $4\sigma$ detection limit of the UDS. As an additional reference, we include the total number counts of the mock redshift survey (black solid line). The empty squares and circles are observations from interferometric and single-dish telescopes, respectively. The solid circles correspond to SPT measurements, known to be dominated by a gravitationally lensed population of galaxies \citep{vieira+2010}. TolTEC/UDS will place strong constrain to the number counts between $S_{\rm 1.1} \sim 0.1\,$ to a few mJy, linking the results from large-area single-dish surveys sensitive to the rare and brighter SMGs down to the fainter population of DSFGs probed by small-area deep interferometic observations.}
    \label{fig:ncounts_toltec_areas}
\end{figure*}

Figure \ref{fig:ncounts_toltec_areas} shows our predictions of the 1.1, 1.4 and 2.0 mm number counts expected to be measured in the UDS/TolTEC survey, and compares them against a compilation of interferometric and single-dish observational results: \citet{vieira+2010,Lindner+2011,scott+2012,karim+2013,hatsukade+2013,staguhn+2014,simpson+2015,fujimoto+2016,hatsukade+2016,geach+2017,franco+2018,magnelli+2019,gonzalez-lopez+2020,zavala+2021,chen+2022,bing+2023}.

Considering a 4$\sigma$ detection limit, the UDS will be sensitive to the faint end ($S_{\rm 1.1} \sim 0.1$\, mJy) of the number counts, probing a regime that has only been sampled by interferometers \citep[e.g.][]{fujimoto+2016,hatsukade+2016,gonzalez-lopez+2020,zavala+2021,chen+2022}. Although these observations benefit from higher angular resolutions, and therefore a reduced confusion noise that makes them sensitive to a fainter population, their relatively small areas ($\lesssim 100$ sq. arcminutes) have limited the studied samples to $\lesssim$ 150 sources. Most of these surveys are therefore potentially affected by sampling variance effects. With a $\sim$ 30 times larger area, the UDS will detect thousands to tens of thousands of DSFGs with $S_{\rm 1.1} = 0.1 - 1.0\,$mJy, allowing to impose strong constrains in this faint regime of the number counts.

On the other hand, although the total area of the UDS will not be able to impose statistically robust constrains on the brightest and more extreme population of SMGs ($S_{\rm 1.1} \gtrsim 10$\, mJy), it will sample and constrain the range of flux densities probed by typical single-dish surveys ($S_{\rm 1.1} \sim 1$\, mJy). Furthermore, the improved angular resolution (FWHM$_{\rm 1.1} \approx 5$ arcsecond) of the TolTEC observations will reduce the impact of source-blending in the measured number counts, an effect that has been confirmed to bias high the measurements from lower angular resolution (FWHM$ > 10$ arcsecond) single-dish observations \citep[e.g.][]{karim+2013, brisbin+2017, Barger+2022}.  
The UDS will therefore provide an important link between the faintest population of DSFGs probed by the deepest small-area interferometric surveys \citep[e.g.][]{fujimoto+2016, gonzalez-lopez+2020}, and the brighter, gravitationally lensed SMGs identified in larger-area but lower angular resolution single-dish observations \citep[particularly at the longer wavelengths, e.g.][]{vieira+2010}.
    
\subsection{The cosmic variance in the UDS/TolTEC surveys}\label{sec:cosmic_var_toltec}

\begin{figure*}
 \centering
  \subfloat{\includegraphics[width=0.38\textwidth]{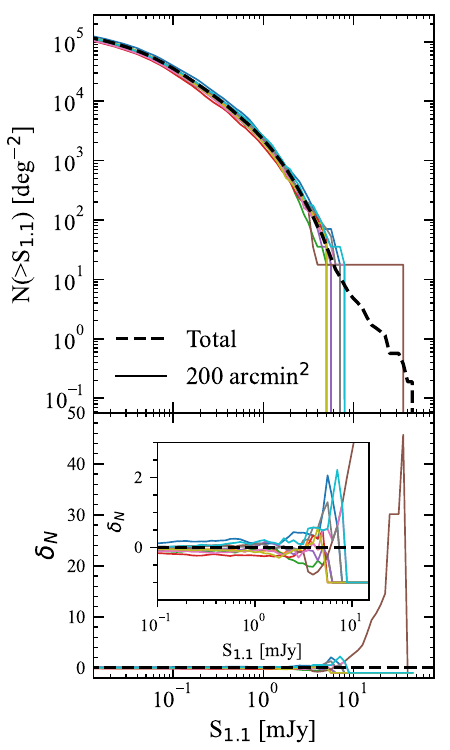}}
  \subfloat{\includegraphics[width=0.379\textwidth]{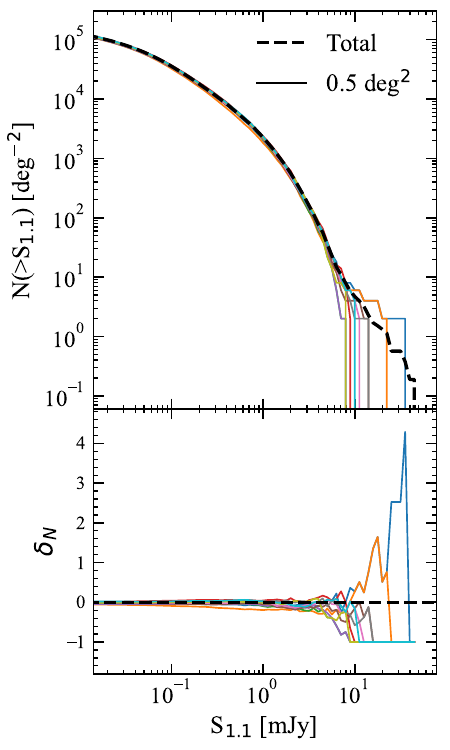}}
 \caption{ Number counts for ten independent areas (colored lines) and their mean number counts for 200 arcmin$^{2}$ and 0.5 deg$^{2}$ (black dashed line in the left and right panels respectively), showing the effect of cosmic variance. The bottom panels show the residual fraction in the number counts obtained by subtracting the mean number counts from each independent realization and normalizing by the mean value.}
 \label{fig:ncounts_toltec_areas_residuo}
\end{figure*}

In Figure \ref{fig:ncounts_toltec_areas_residuo} we present the number counts at 1.1 mm for ten independent areas of 200 arcmin$^{2}$ and 0.5 deg$^{2}$ (representative of the smaller and larger areas considered for the UDS) along with the number counts for the total area of the mock redshift survey. We also add the residual fraction ($\delta_{\rm N}$) obtained by subtracting the total number counts from each realization and normalizing by the total. Our mock surveys predict that, for the smaller 200 arcmin$^{2}$ maps, cosmic variance can introduce small relative variations of $\delta_{\rm N} < $\,0.3 at $S_{\rm 1.1} < 2\,$mJy, systematically increasing with flux density and reaching $\delta_{\rm N} \sim $\,2 at $2\;\rm{mJy}< S_{\rm 1.1} < 7\,$mJy. For higher flux densities the variation can be more significant, $\delta_{\rm N} \sim$\,45. In the case of the 0.5 sq. degree map, the cosmic variance starts to noticeable affect the counts ($\delta_{\rm N} \sim $\,1) at $\sim$10 mJy, reaching $\delta_{\rm N} \sim $\,4.3 for the highest flux.

\subsection{UDS/TolTEC Luminosity Function}\label{sec:toltec_lf}

\begin{figure*}
\makebox[\textwidth]{\includegraphics[width=\textwidth]{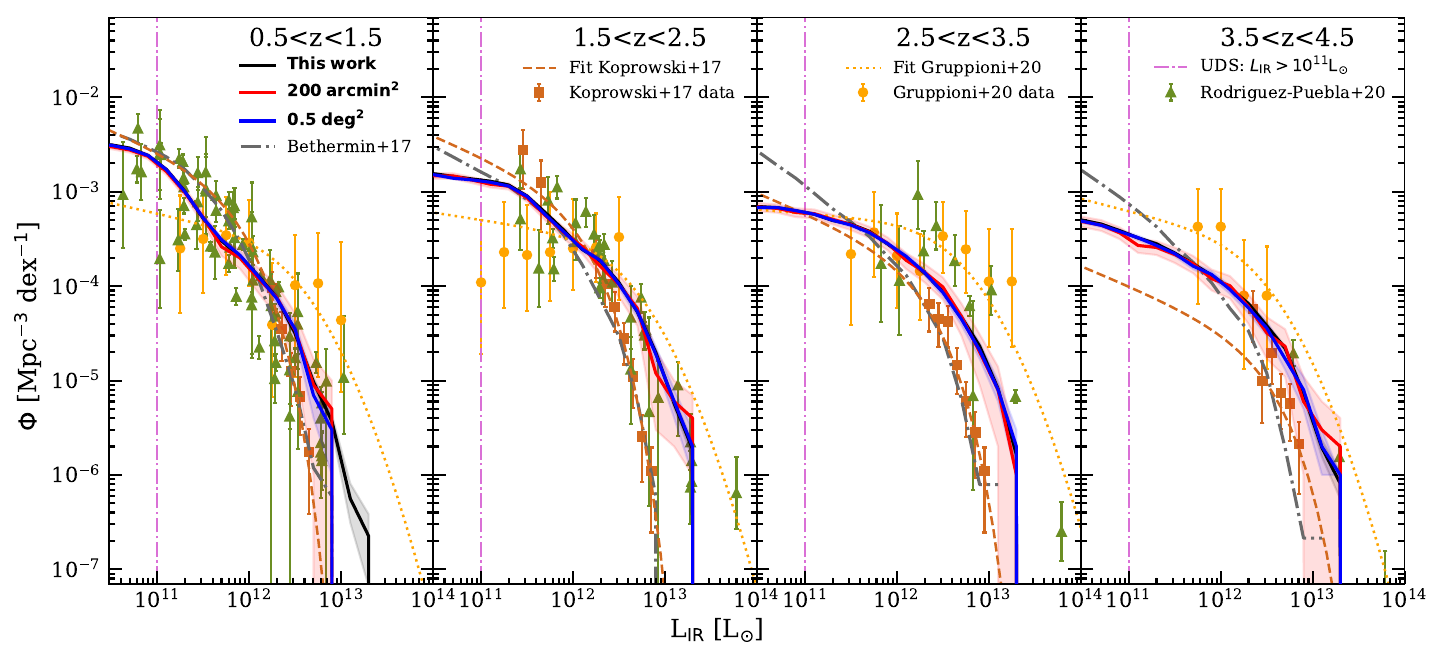}}
\caption{Our predictions of the mean infrared luminosity function measured for the UDS/TolTEC areas (red and blue solid lines) spanning from $z=0.5$ to $z=4.5$. The colored shadow areas correspond to the $1\sigma$ error for the mean. The results for 5.3deg$^{2}$ is also presented (black solid line). The different areas with different depths allow us to study the luminosity function in a wide interval of luminosity. We compare our luminosity function with previous results \citep{koprowski+2017, gruppioni+2020, R-P+2020} mark as different colored symbols (brown squares, yellow circles and green triangles). The discontinuous lines with the same colors than the symbols are fits to the respective data. The vertical line indicates the detection threshold in luminosity for the UDS/TolTEC survey.}
    \label{fig:lf_toltec_areas}
\end{figure*}

Similar to the number counts, the different areas and depths of the TolTEC extragalactic legacy surveys will allow us to study different regions of the luminosity function. In particular, the UDS has an infrared luminosity lower limit of $\sim 10^{11}$ L$_{\sun}$, enabling the detection of Luminous Infrared Galaxies (LIRGs), with SFRs of  dozens of M$_{\sun}$ yr$^{-1}$. Figure \ref{fig:lf_toltec_areas} shows our predictions of the luminosity function we expect to be sensitive to 
in the two characteristic areas of the UDS. The results are divided in four redshift bins, spanning from $z=0.5$ to $z=4.5$, and are compared against the same data sets included in Figure \ref{fig:lf}. The UDS will therefore constrain very well (i.e. with little cosmic variance) the shape of the luminosity function in the less luminous regime, addressing some discrepancies noted in previous studies \citep{koprowski+2017, gruppioni+2020, R-P+2020}. At $L_{\rm{IR}} > 4\times10^{12}\; \rm{L}_{\sun}$, the 200 arcmin$^{2}$ area exhibits more variance than the 0.5 deg$^{2}$ map, as it becomes challenging to capture brighter galaxies. Nevertheless, the 0.5 deg$^{2}$ area will provide more reliable luminosity function estimates up to $\sim L_{\rm{IR}} \sim 10^{13}\; \rm{L}_{\sun}$. While the UDS survey will contribute to probe part of this brighter end of the LF, its capabilities remain limited in this regard. More precise insights into this region of the luminosity function will be obtained with the shallower ($1\sigma_{\rm 1.1} \sim 0.25$\, mJy) but $\sim 75$ times wider LSS/TolTEC surveys.

\subsection{UDS/TolTEC redshift distribution}
\label{toltec_zdist}

\begin{figure}
\centering
\includegraphics[width=\columnwidth]{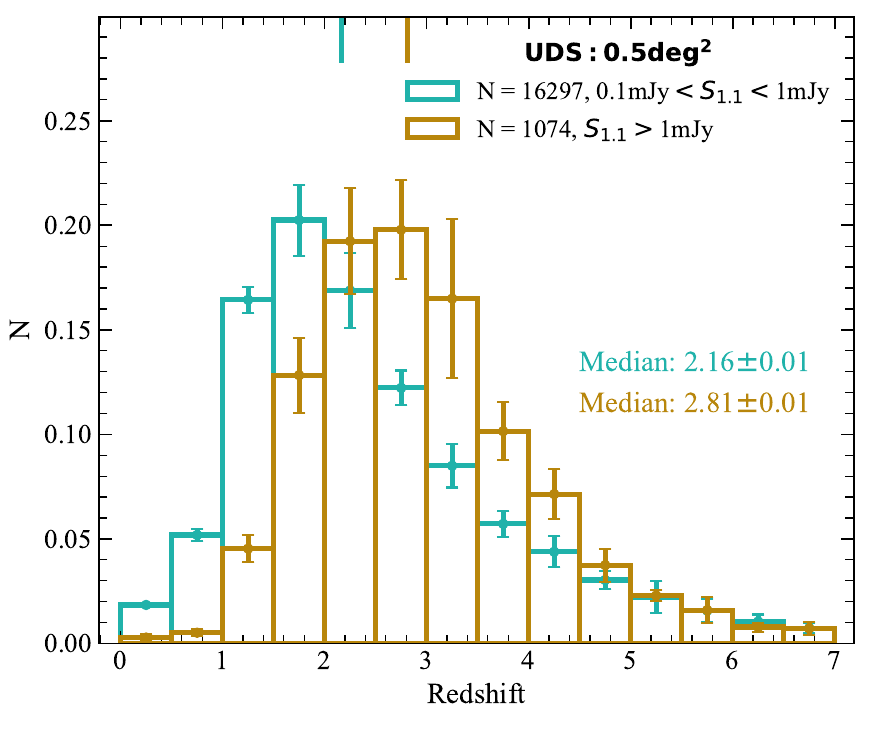}
\caption{Redshift distribution at 1.1 mm for selected DSFGs in our mock redshift survey in an area of 0.5 sq. degree to be observed as part of the UDS/TolTEC survey. The cyan and golden distributions correspond to galaxies with flux densities of $0.1 <S_{1.1} [{\rm mJy}] < 1.0$ mJy and $S_{\rm 1.1} > 1$ mJy respectively. The errors bars represent the 1$\sigma$ dispersion from the ten 0.5 sq. degree map realizations. The median values for each distribution are indicated by the upper vertical lines. The difference between the bright and faint DSFG redshift distribution can be attributed to the known downsizing effect. }
    \label{fig:zdist_toltec_areas}
\end{figure}

Our predictions of the DSFG redshift distribution expected in the larger 0.5 sq. degree area of the UDS are presented in Figure \ref{fig:zdist_toltec_areas}. To explore the dependence of the redshift distribution with flux density, the population of DSFGs in our mock redshift survey is divided in two flux density intervals, a fainter sample with $0.1 < S_{\rm 1.1} {\rm [mJy]} < 1.0$ and a brighter sample with $S_{1.1} \geq 1.0\,\rm{mJy}$. A clear difference between the redshift distributions can be seen, with the fainter population dominating at lower redshifts ($z_{\rm med}=2.16\pm0.01$) and brighter sources being more abundant at higher redshifts ($z_{\rm med}=2.81\pm0.01$). This behavior, which is present in some of the observational results discussed in \S \ref{subsec3:zdist}, can be attributed to the known cosmic downsizing. That is, the fact that more massive galaxies form earlier and faster than less massive ones \citep[see e.g.,][]{Firmani+2010}.

The area and depth of the UDS, combined with the $\sim5$\, arcsecond angular resolution of TolTEC on the 50-LMT and deep multi-wavelength data available in the targeted fields, will result in a large sample of $\sim 10,000$ galaxies with robust Optical/IR counterparts and photometric redshifts. This will allow us to measure the redshift distribution of DSFGs, and study its dependence with different physical properties, at a level not yet possible with current data sets.

%------------------SECTION 5-------------------

\section{The Impact of galaxy clustering on Flux boosting}\label{sec4:boosting_clustering}

The measured flux densities of sources in a real observation can sometimes be overestimated due to several factors, including the contribution from faint unresolved sources and the influence of neighbouring sources that may be blended together within the telescope beam. This effect, commonly referred to as `flux boosting', is known to be stronger in sources detected at low signal-to-noise ratio (S/N), as discussed in previous studies \citep[e.g.][]{hogg_+1998}. Most studies rely on simulations to estimate and correct for flux boosting effects \citep[e.g.][]{zavala+2017,geach+2017,stach+19}. Most of these simulations, however, randomly distribute sources on noise maps without taking into account any clustering information, which could potentially bias the flux boosting of sources.

Taking advantage of the clustering information in our mock redshift survey, we investigate its impact on the estimation of flux boosting. We measure the effect of flux boosting at the three TolTEC wavelengths (1.1, 1.4 and 2.0 mm), adopting the corresponding angular resolution (FWHM = 5.0, 6.3 and 9.5 arcsec) and expected r.m.s. noise level ($\sigma_{\rm r.m.s.} = $ 0.025, 0.018, and 0.012 mJy) of the UDS survey. We then compare our results to a mock redshift survey with randomly distributed sources.

We categorize the galaxies in our mock redshift survey into two distinct groups. The first group, referred to as `candidates', comprises galaxies with flux densities exceeding the r.m.s. noise level in the UDS (i.e., $S_{1.1}>0.025$ mJy, $S_{1.4}>0.018$ mJy, $S_{2.0}>0.012$ mJy). Fainter galaxies, unlikely to be individually detected but still contributing to the background noise, are denoted as `boosting sources' and are included in the second group.

\begin{figure*}
  \centering
    \subfloat{\includegraphics[width=1\textwidth]{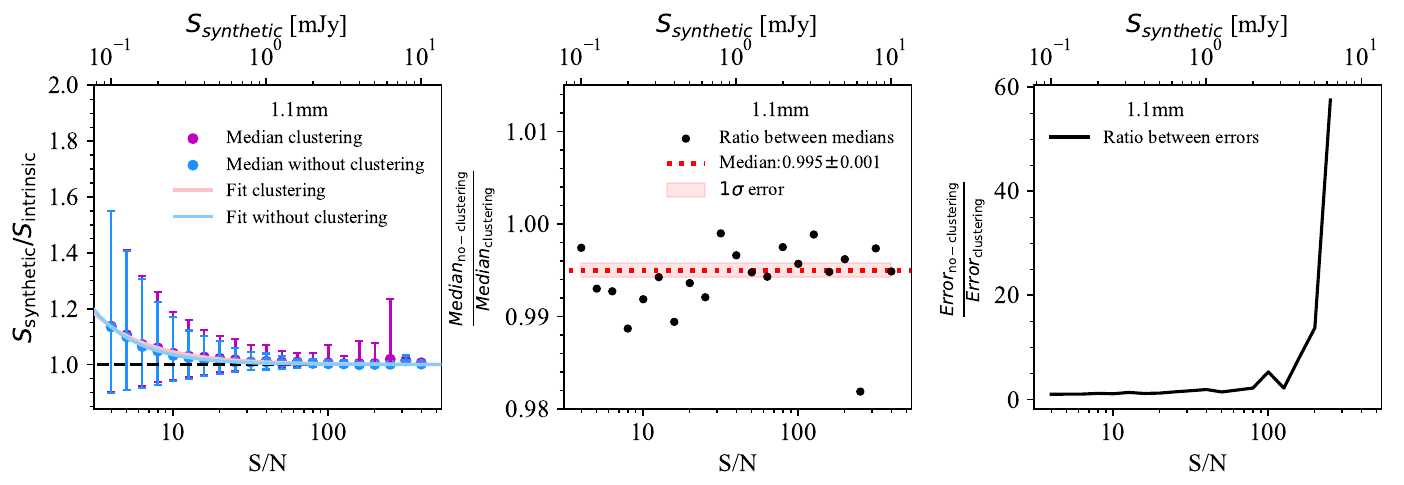}}

    \subfloat{\includegraphics[width=1\textwidth]{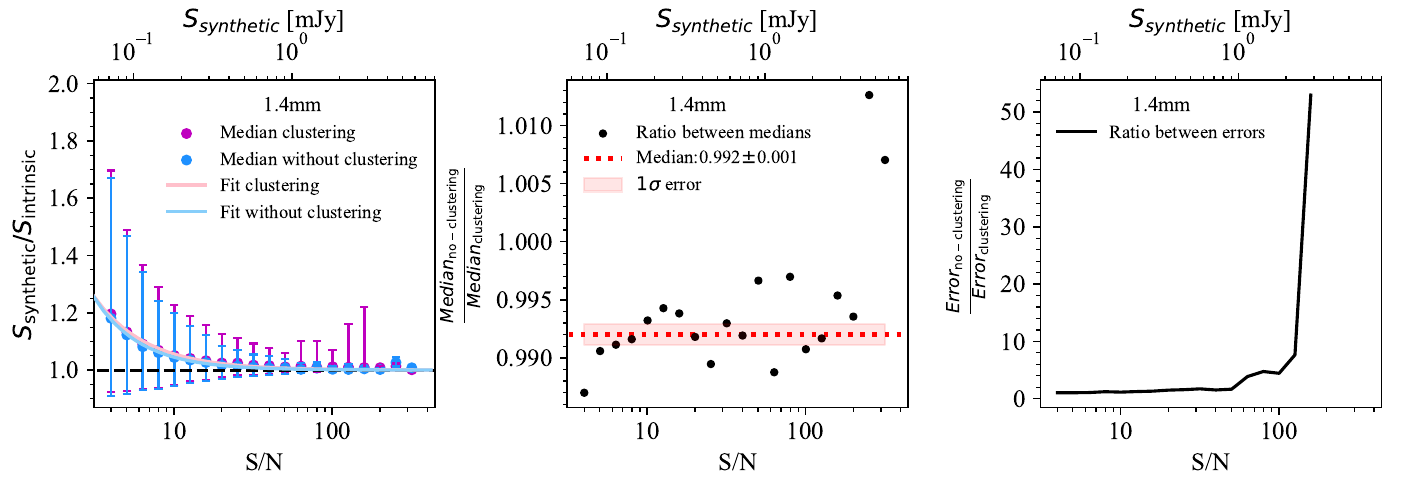}}

    \subfloat{\includegraphics[width=1\textwidth]{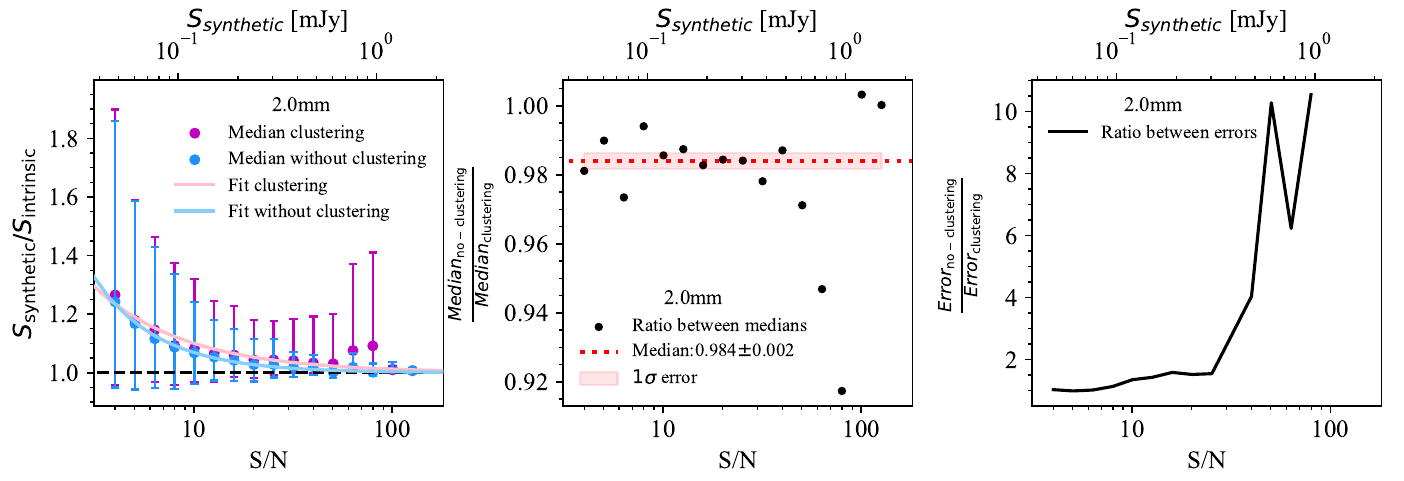}}
\caption{{\it Left:} Median boosting factor vs signal-to-noise ratio for an area of 1 deg$^{2}$ obtained at 1.1, 1.4 and 2.0 mm (top, middle, and bottom panels). The magenta dots represent the median boosting factor measured in a map including galaxy clustering, while the blue dots correspond to the median boosting factor when sources are assigned random positions. The pink and blue solid curves are a fits to the data with and without clustering. The vertical bars represent the $1\sigma$ error in the medians. {\it Middle:} ratio between the median boosting factor values without and with clustering of the DSFG population. The red horizontal lines represent the median value of the ratios, with the shaded area representing the $1\sigma$ error of the median. {\it Right:} ratio between positive errors in the median boosting factor.}
    \label{fig:boosting_hr}
\end{figure*}

Each candidate source, with intrinsic flux density $S_{\rm instrinsic}$, is convolved with a Gaussian telescope beam in order to incorporate the flux density contribution from surrounding fainter sources. The resulting flux density is denoted as $S_{\rm{blend}}$. The noise contribution ($S_{\rm noise}$) at each source position is drawn from a Gaussian distribution with $\sigma_{\rm r.m.s.}$ depending on the specific observing wavelength. The resulting `synthetic' flux density of each source is given by $S_{\rm{synthetic}} = S_{\rm{blend}} + S_{\rm{noise}}$. Finally, the boosting factor is defined as the ratio between the synthetic and the intrinsic flux densities ($S_{\rm{synthetic}}/S_{\rm{intrinsic}}$). The same methodology is applied to the mock redshift survey with the galaxies randomly distributed (i.e. removing clustering information). Figure \ref{fig:boosting_hr} shows the median boosting factor, with $1\sigma$ errors estimated through a bootstraping analysis, as a function of synthetic flux density ($S_{\rm{synthetic}}$) and signal-to-noise ration (S/N). Additionally to model the behavior of the boosting factor in the mock redshift survey we fit to our data a power law following \citet{geach+2017} with the form:

\begin{equation}
    B({\rm S/N}) = 1+a \left ( \frac{{\rm S/N}}{b} \right ) ^{-c}
    \label{ec:fit_boosting}
\end{equation}

\noindent with the fitted parameters for each data set given in Table \ref{tab:fit_boosting}.

\begin{table}
    \caption{Parameters fitted to equation (\ref{ec:fit_boosting}) to model the boosting factor in maps with and without clustering.}
    \label{tab:fit_boosting}
    \centering
    \begin{tabular}{lccc}
    \hline
     \multicolumn{1}{c}{} & $a$ & $b$ & $c$  \\ \hline
     \multicolumn{1}{c}{} & \multicolumn{3}{c}{1.1 mm} \\
     Clustered & 2.04 & 0.4 & 1.18 \\ 
     Random & 1.6 & 0.73 & 1.46 \\
     \multicolumn{1}{c}{} & \multicolumn{3}{c}{1.4 mm} \\
     Clustered & 0.65 & 1.53 & 1.3 \\
     Random  & 1.6 & 0.89 & 1.49 \\
     \multicolumn{1}{c}{} & \multicolumn{3}{c}{2.0 mm} \\
     Clustered & 0.65 & 1.32 & 0.92\\
     Random & 1.6 & 0.96 & 1.34\\ \hline
    \end{tabular}

\end{table}

As expected, fainter galaxies (i.e. S/N $\sim 4$) are more likely to be strongly boosted by unresolved neighbour sources, with median boosting factors of $1.14_{-0.23}^{+0.41}$, $1.20_{-0.27}^{+0.50}$ and $1.26_{-0.30}^{+0.63}$ for the clustering map for the three wavelengths. The median values of the boosting factors are very similar for both clustered and non-clustered populations at 1.1 mm and 1.4 mm. At 2.0 mm, we observe departures at S/N $>7$. The larger beam size allows for a larger chance of blending sources. This is the dominant effect in the abrupt departures from the non-clustering solution seen at S/N $>60$ (i.e. $S_{\rm 2.0} >  0.72\,$ mJy, see Figure \ref{fig:boosting_hr}), where several sources of similar brightness are blended within the same beam. At the largest S/N there are no differences in the median values, since the surface density of bright galaxies is smaller. 

The errors of the median values of the boosting factors are systematically larger when clustering is considered, specially for the larger S/N values. This indicates the high uncertainty to correct for boosting individual high S/N sources. This is in contrast with the findings of previous studies \citep{zavala+2017, geach+2017, stach+19}, where only faint galaxies were observed to be significantly affected by boosting in maps without clustering.

To quantify our results, Figure \ref{fig:boosting_hr} shows the ratio between the boosting factors estimated for a non-clustered and a clustered population of DSFGs. On average, clustering increases the boosting factor by 0.5$\pm$0.1 per cent at 1.1 mm, 0.8$\pm$0.1 per cent at 1.4 mm, and 1.6$\pm0.2$ per cent at 2.0 mm. Despite being a relatively small correction, including clustering in the flux boosting determinations will systematically improve the flux density measurements, particularly for the lower angular resolution observations.

%-----------------------SECTION 6----------------------

\section{Summary and Conclusions}\label{sec5:discussion_conclution}

In this paper, we present a new cosmological mock redshift survey of the DSFG population based on a lightcone built from the Bolshoi-\textit{Planck} dark matter halo simulation \citep{klypin+2016,RP+2016b}. The mock redshift survey covers a total area of 5.3 sq. degree and a redshift range $z=0-7$. The merger trees of dark matter haloes are used in order to model the mass assembly and star formation histories of galaxies based on the updated version of semi-empirical model by \citet{RP+2017}. We used a sigma clipping process to define a star forming main sequence from our mock survey and found that it reproduces previous determinations from the literature \citep{Speagel+2014,popesso+2022,cardona-torres+2022}. The infrared properties of the galaxies are assigned based on empirical determinations of the dust-obscured fraction of SFR \citep{whitaker+2017,R-P+2020} and the SFRD \citep{dunlop+2017}, the \citet{kennicutt_1998} relation between SFR and $L_\text{IR}$, the relation between the observed wavelength peak, $\lambda_{\rm peak}$, of the IR SED and $L_{\rm IR}$ \citep{casey+2018a} and the Wien's displacement law to obtain the dust temperature (along with the assumption of modified black body for the SED), which are corrected by heating effects due to the CMB \citep{dacunha+2013}. The alignment between dark matter haloes and background galaxies is used to incorporate gravitational lensing magnifications \citep{narayan+1995}. 

This new mock redshift survey accurately reproduces the mm wavelength number counts, from the faint end ($S_{\rm 1.1} \sim 0.1\,$ mJy) probed by small area interferometric observations to the bright end ($S_{\rm 1.1} \sim 10\,$ mJy), mostly explored by larger single-dish surveys. Although the total area of the mock redshift survey is not large enough to include a large sample of the brighter and more extreme SMGs ($S_{\rm 1.1} > 10.0\,$ mJy), it does reproduce the expected flattening (at $S_{\rm 1.1} \gtrsim 8\,$ mJy) due to the strongly gravitationally lensed population of galaxies. 
Furthermore, although the evolution of the IR luminosity functions is still not well constrained by observations, particularly at $z > 2.5$, our mock redshift survey predicts a luminosity function that falls well within the uncertainties and dispersion of the different observational results, from $z = 0.5$ to 4.5 and over two orders of magnitude (${\rm log}[L_{\rm IR}/{\rm L}_\odot] = 11 - 13$). 

Additionally, our mock redshift survey reproduces the total and IR measurements of the SFR density from $z = 0-7$. Compared to recent studies that have estimated the redshift distribution of faint DSFGs ($S_{\rm 1.1} \gtrsim 0.2\,$ mJy), our mock redshift survey systematically predicts slightly ($\lesssim 10\,$ per\,cent) higher median redshift values, suggesting that we might be missing a small fraction of faint low-$z$ DSFGs or overestimating the high-$z$ population. It should be noted, however, that most of these observational results are based on relatively small (< tens of sq. arcminutes) interferometric surveys, and can therefore be affected by cosmic variance effects. Our mock redshift survey further predicts the effect of cosmic downsizing, with brighter DSFGs being more abundant in the past and a fainter population dominating at lower redshifts. 

It should be noted that the predictions of our mock redshift survey for the fainter population of DSFGs (i.e. $S_{\rm 1.1} \lesssim 0.05\,$mJy and $L_{\rm IR} \lesssim 10^{11} {\rm L}_\odot$) are potentially underestimated and limited by the mass resolution of the dark matter cosmological simulation ($1.55\times 10^{8}$ $h^{-1}$M$_{\sun}$). Similarly, the brighter population (i.e. $S_{\rm 1.1} \gtrsim 10\,$mJy and $L_{\rm IR} \gtrsim 10^{13} {\rm L}_\odot$) is poorly sampled by the total area of the survey (5.3 sq. degree). A larger mock redshift survey ($A \sim 100\,$sq. degree) to probe this brighter population and its clustering properties is under development and will be presented in a forthcoming paper.

Mock redshift surveys like the one presented in this work, are an essential tool for the design of future extragalactic surveys and to predict the different properties of the population of galaxies that they will probe. Furthermore, they can be used to develop and test source extraction techniques and data analysis tools, characterize different biases affecting real observations, estimate the expected confusion noise, between others. In this work we take advantage of our mock redshift survey to make predictions for the TolTEC Ultra Deep Survey ($A \approx 0.8\,$ sq. degree, $1\sigma_{\rm 1.1} \approx 0.025\,$mJy) and estimate the impact of clustering in flux boosting determinations. Below, we summarize our  main results regarding these questions:

\begin{itemize}
\item The UDS will detect $\sim 24,000$ DSFGs above a $4\sigma$ threshold (i.e. $S_{\rm 1.1} \gtrsim 0.1\,$ mJy). This sample will result in a comprehensive view of the 1.1, 1.4, and 2.0 mm number counts between $S_{\rm 1.1} \sim 0.1\,$ and 30 mJy. The $S_{\rm 1.1} \sim 0.1 - 1.0\,$ mJy regime will be strongly constrained, providing a robust connection between deep interferometric observations and the wider but shallower single-dish surveys. Towards the brighter end, cosmic variance will have a $\sim 20\,$per\,cent impact on the counts at $S_{\rm 1.1} \sim 5\,$mJy, increasing to $> 40\,$per\,cent for $S_{\rm 1.1} > 10\,$mJy.

\item The UDS sample will be dominated by relatively low mass DSFGs with $M_\star = 10^{9.5} - 10^{10.5}\, {\rm M}_\odot$, with $\sim 17\,$per\,cent of the sample corresponding to $M_\star < 10^{9.5}\, {\rm M}_\odot$ galaxies. This will allow us to explore a population an order of magnitude less massive than the typical SMGs identified with previous single-dish surveys.

\item The area and sensitivity of the UDS, combined with the $\sim5$\, arcsecond angular resolution of TolTEC on the 50-LMT and deep multi-wavelength data available in the UDS fields, will result in a large sample of DSFGs with robust Optical/IR counterparts. This will allow us to estimate accurate physical properties and photometric redshifts, and study the evolution of the IR luminosity function in a wide redshift range ($0.5 < z < 4.5$), particularly in the LIRG regime ($10^{11} < L_{\rm IR}/{\rm L}_\odot < 10^{12}$). Furthermore, the UDS will constrain the redshift distribution of relatively faint DSFGs and its dependence with different physical properties. This will provide important insights on the effect of downsizing in this population of galaxies.

\item The median flux boosting of sources detected with a S/N = 4 in the UDS will be $1.14_{-0.23}^{+0.41}$, $1.20_{-0.27}^{+0.50}$ and $1.26_{-0.30}^{+0.63}$ at 1.1, 1.4, and 2.0 mm. As expected, flux boosting has a smaller impact on brighter sources detected with a higher signal-to-noise, e.g. for $10\sigma$ sources, the mean boosting factors and their dispersion are reduced to $1.04_{-0.10}^{+0.15}$, $1.05_{-0.10}^{+0.18}$ and $1.08_{-0.12}^{+0.23}$.

\item Including clustering has a relatively minor impact on the determination of flux boosting factors, with no evident dependence on source S/N. Compared to a randomly distributed population, clustering increases flux boosting by 0.5$\pm$0.1, 0.8$\pm$0.1 and 1.6$\pm$2.0\,per\,cent at 1.1, 1.4, and 2.0 mm. The relative increase with angular resolution suggests that clustering might have a stronger impact on lower angular resolution observations (e.g. FWHM$\gtrsim 10$\,arcsec). Clustering, however, increases the uncertainties associated to the flux boosting factor determination, and should be propagated to the flux density uncertainties.

\end{itemize}

%--------------------ACKNOWLEDGEMENTS--------------------------------
\newpage
\section*{Acknowledgements}

We thank the anonymous referee for the constructive comments and observations that have improved this work. This research has been possible thanks to the Mexican PhD scholarship supported by the Consejo Nacional de Humanidades Ciencias y Tecnologías (CONAHCyT). This work has been supported by CONAHCYT through projects: CB 2016 - 281948, A1-S-45680 and ``Ciencia de Frontera'' G-543. VAR acknowledges partial support from grant PAPIIT-UNAM IN106823. ARP acknowledges financial support from CONACyT ``Ciencia Basica'' grant 285721 and DGAPA-PAPIIT IN106124.

%---------------DATA AVAILABILITY-----------------------

\section*{Data Availability}

The mock redshift survey will be available for download at 
\href{https://mnemosyne.inaoep.mx/index.php/s/phXy5KLx09FumYw}{https://mnemosyne.inaoep.mx/index.php/s/phXy5KLx09FumYw} or on request from the corresponding author.

%%%%%%%%%%%%%%%%%%%% REFERENCES %%%%%%%%%%%%%%%%%%

% The best way to enter references is to use BibTeX:

\bibliographystyle{mnras}
\bibliography{bibliography} % if your bibtex file is called example.bib

% Alternatively you could enter them by hand, like this:
% This method is tedious and prone to error if you have lots of references
%\begin{thebibliography}{99}
%\bibitem[\protect\citeauthoryear{Author}{2012}]{Author2012}
%Author A.~N., 2013, Journal of Improbable Astronomy, 1, 1
%\bibitem[\protect\citeauthoryear{Others}{2013}]{Others2013}
%Others S., 2012, Journal of Interesting Stuff, 17, 198
%\end{thebibliography}

%%%%%%%%%%%%%%%%%%%%%%%%%%%%%%%%%%%%%%%%%%%%%%%%%%

%%%%%%%%%%%%%%%%% APPENDICES %%%%%%%%%%%%%%%%%%%%%

\appendix

\section{Comparison with SIDES}
In the Table \ref{tab:appendix_A} we summarize the main differences between our mock redshift survey and that of SIDES  \citep{bethermin+2017}.

\begin{table*}
\caption{Main differences between the prescription to generate our mock redshift survey and that from SIDES \citep{bethermin+2017}.}
\label{tab:appendix_A}
 \begin{tabular}{p{4cm} p{6cm} p{6cm}}
\toprule
\multicolumn{1}{c}{Main differences} &
  \multicolumn{1}{c}{This Work} &
  \multicolumn{1}{c}{SIDES} \\ \midrule
Area &
  5.3 sq. degree &
  2 sq. degree \\ \midrule
Star-Forming galaxy selection &
  We use the criteria established by \citet{pacifici+2016} to preselect star-forming galaxies and separate them into redshift and stellar mass bin. We then calculate the mean sSFR value in each stellar mass bin and perform a linear fit. The galaxies above the fit minus 0.5 dex are selected as star-forming. This process is repeated iteratively until the fit converges. &
  The authors estimate the probability of galaxies being star-forming according to their $M_{\star}$ and $z$. The probability is a fit of the observed evolution of the star-forming galaxy fraction by \citet{davidzon+2017}. \\ \midrule
Determination of SFRs &
  We use the model develop by \citet{RP+2017}, which follows the growth in mass of the dark matter haloes through time to predict the $M_{\star}$ as a function of $z$ and, therefore, their history of SFR. The model takes into account the fraction of star-formation \textit{in-situ} and \textit{ex-situ} (i.e. mergers). The SFR is separated into its UV and IR contributions. The $L_{\rm{IR}}$ is only given by the IR component. We impose an upper limit to the SFR$_{\rm{IR}}$ of 2000 M$_{\sun}$ yr$^{-1}$. &
 The authors apply the equation by \citet{schreiber+15} that measures the evolution of the main sequence of star-forming galaxies according their $z$ and $M_{\star}$. Starburst galaxies are randomly introduced, following a fraction of 1.5 percent at $z=0$, increasing to 3 per cent at $z=1$, and then remaining constant. The maximum total SFR allowed is 1000 M$_{\sun}$ yr$^{-1}$ and it is directly converted to $L_{\rm{IR}}$. \\ \midrule
Dust temperature, $T_{\rm{d}}$ &
  $T_{\rm{d}}$ is estimated for each galaxy using an approximation to Wien's displacement law, with the peak wavelength of the SED given by the $\lambda_{\rm peak}-L_{\rm{IR}}$ relation from \citet{casey+2018a}. These $T_{\rm{d}}$ values are then corrected for CMB heating effects following \citet{dacunha+2013}. &
  The dust temperatures are $T_{\rm{d}}\sim 40$ K defined by the SED assigned to each galaxy, and no CMB heating corrections are applied. \\ \midrule
Observed flux densities, $S_{\nu}$ &
  We assume that the emission of our galaxies is well modeled by a modified black body, determined by the dust temperature of our galaxies, their $L_{\rm{IR}}$ and an emissivity index of $\beta=1.8$. &
  Flux densities are determined using SEDs from the \citet{magdis+2012} library, which depends on the mean intensity of the radiation field that have different values whether the galaxy is star-forming or starburst. Only the star-forming galaxies selected have far-infrared and millimetre properties. \\ \midrule
Gravitational lensing effect &
  We consider strong amplifications ($\mu>2$) produced by dark-matter haloes on background galaxies using a simplified point-mass model \citep{narayan+1995}. For each halo in our mock redshift survey, nearby galaxies within a given radius are amplified based on the halo mass and the position of the source respect the line of sight and its distance. &
  Lensing magnifications are randomly assigned (i.e. not considering spatial information). Strong magnifications ($\mu>2$) are introduced using the redshift-dependent probability distribution of \citet{hezaveh+11}, while weak lensing magnifications are drawn from a Gaussian distribution with parameters from \citet{hilbert+07}. \\ \bottomrule
\end{tabular}
\end{table*}

%%%%%%%%%%%%%%%%%%%%%%%%%%%%%%%%%%%%%%%%%%%%%%%%%%

% Don't change these lines
\bsp	% typesetting comment
\label{lastpage}
\end{document}